\RequirePackage{fix-cm}
\RequirePackage{amsmath}
\RequirePackage{physics}
\documentclass[twocolumn]{svjour3}
\smartqed  
\RequirePackage{graphicx}
\RequirePackage{placeins}
\RequirePackage[utf8]{inputenc} 
\RequirePackage[numbers,sort&compress]{natbib}
\newcommand{\dyturbo}{\texttt{DYTurbo}}
\newcommand{\dyres}{\texttt{DYRes}}
\newcommand{\dyqt}{\texttt{DYqT}}
\newcommand{\dynnlo}{\texttt{DYNNLO}}

\newcommand{\vegas}{Vegas}
\newcommand{\quadra}{Quadrature}
\newcommand*{\Zboson}{\ensuremath{Z}}
\newcommand*{\Vboson}{\ensuremath{V}}

\newcommand*{\Wboson}{\ensuremath{W}}
\newcommand*{\Zgamma}{\ensuremath{Z/\gamma^*}}

\newcommand*{\pT}{\ensuremath{p_{\mathrm{T}}}}

\newcommand*{\qT}{\ensuremath{q_{\mathrm{T}}}}
\newcommand*{\qt}{\ensuremath{q_{\mathrm{T}}}}

\newcommand*{\QT}{\ensuremath{Q_{\mathrm{T}}}}
 
\newcommand\as{\ensuremath{\alpha_{\mathrm{S}}}} 
\journalname{Eur. Phys. J. C}
\usepackage{subfigure}
\usepackage{booktabs}
\usepackage{epstopdf}
\RequirePackage[colorlinks,citecolor=blue,urlcolor=blue,linkcolor=blue]{hyperref}
\usepackage{url}
\RequirePackage[T1]{fontenc}
\usepackage[british]{babel}
\usepackage{microtype}
\begin{document}
\title{DYTurbo: Fast predictions for Drell--Yan processes}
\titlerunning{\dyturbo}        
\author{Stefano Camarda$^{*}$$^{1}$, Maarten Boonekamp$^{2}$, Giuseppe
  Bozzi$^{3}$, Stefano Catani$^{4}$, Leandro Cieri$^{5}$, Jakub
  Cuth$^{6}$, Giancarlo Ferrera$^{7}$, Daniel de Florian$^{8}$, Alexandre Glazov$^{9}$,
  Massimiliano Grazzini$^{10}$, Manuella G. Vincter$^{11}$, Matthias
  Schott$^{6}$
}
\institute{
  \at $^{1}$CERN, Geneva, Switzerland
  \at $^{2}$SACLAY, Paris, France
  \at $^{3}$Dipartimento di Fisica, Universit\`a di Pavia and INFN, Sezione di Pavia, Italy
  \at $^{4}$INFN, Sezione di Firenze, Italy
  \at $^{5}$INFN, Sezione di Milano-Bicocca, Italy
  \at $^{6}$Johannes Gutenberg-University, Mainz, Germany
  \at $^{7}$Dipartimento di Fisica, Universit\`a di Milano and INFN, Sezione di Milano, Italy
  \at $^{8}$ICAS-UNSAM, San Martin, Argentina
  \at $^{9}$DESY, Hamburg, Germany
  \at $^{10}$Department of Physics, University of Zurich, Switzerland
  \at $^{11}$Carleton University, Ottawa, Canada
  \at $^{*}$Corresponding author
}
\authorrunning{S.~Camarda et al.} 
\date{}
\maketitle
\begin{abstract}
  Drell--Yan lepton pair production processes are extremely important
  for Standard Model (SM) precision tests and for beyond the SM
  searches at hadron colliders. Fast and accurate predictions are
  essential to enable the best use of the precision measurements of
  these processes; they are used for parton density fits, for the
  extraction of fundamental parameters of the SM, and for the
  estimation of background processes in searches. This paper describes
  a new numerical program, \dyturbo, for the calculation of the QCD
  transverse-momentum resummation of Drell--Yan cross sections up to
  next-to-next-to-leading logarithmic accuracy combined with the
  fixed-order results at next-to-next-to-leading order
  ($\mathcal{O}(\as^2)$), including the full kinematical
  dependence of the decaying lepton pair with the corresponding spin
  correlations and the finite-width effects. The \dyturbo{} program is
  an improved reimplementation of the \dyqt, \dyres{} and \dynnlo{}
  programs, which provides fast and numerically precise predictions
  through the factorisation of the cross section into production and
  decay variables, and the usage of quadrature rules based on
  interpolating functions for the integration over kinematic
  variables.
\end{abstract}
\section{Introduction}
\label{intro}
The Drell--Yan process denotes massive lepton-pair production in
hadron-hadron collisions at high energies, as proposed by Sidney
D. Drell and Tung-Mow Yan in 1970~\cite{PhysRevLett.25.902.2}, and
first observed at the Alternating Gradient
Synchrotron~\cite{PhysRevLett.25.1523}.
At the Large Hadron Collider (LHC)~\cite{1748-0221-3-08-S08001}, the
Drell--Yan process continues to play a fundamental role in probing the
proton parton distribution functions (PDF), thereby providing
valuable information on the $u$- and $d$-quark valence
PDFs~\cite{Khachatryan:2016pev} and insight into the
light-quark sea decomposition, in particular on the $s$- over
$\bar{d}$-quark ratio~\cite{Aaboud:2016btc}. This process is also used
to measure fundamental electroweak parameters such as the mass of the
\Wboson\ boson~\cite{Aaboud:2017svj}, the weak-mixing
angle~\cite{Aad:2015uau,Chatrchyan:2011ya}, and the \Wboson-boson
width~\cite{Camarda:2016twt}. An accurate modelling of the Drell--Yan
process is of paramount importance for searches of new physics
phenomena beyond the Standard Model (SM) in final states with high
dilepton invariant
mass~\cite{Khachatryan:2016zqb,Khachatryan:2016jww,Aaboud:2016cth,Aaboud:2016zkn}.
These experimental measurements need to be compared to accurate
predictions based on high-order perturbative QCD and electroweak
corrections.
The Drell-Yan production total cross section and the vector boson
rapidity distribution have been analytically computed up to the
next-to-next-to-leading order (NNLO) in 
powers of the QCD coupling  $\alpha_{\mathrm{S}}$
in Refs.~\cite{Hamberg:1990np,Harlander:2002wh}
and \cite{Anastasiou:2003ds}, respectively.
Fully exclusive parton-level NNLO calculations, which include the
leptonic decay of the vector boson, have been implemented in publicly
available Monte Carlo
codes~\cite{Melnikov:2006di,Melnikov:2006kv,Catani:2009sm,Catani:2010en}.
The transverse-mo\-men\-tum (\qT{}) distribution of the lepton pair
at large (formally, non-vanishing) values of \qT{} can be evaluated at 
$O(\alpha_{\mathrm{S}}^3)$ from the parton-level calculations of \Wboson/\Zgamma + jet production that have been performed in 
Refs.~\cite{Boughezal:2015dva,Ridder:2015dxa,Boughezal:2015ded,Boughezal:2016dtm,Gehrmann-DeRidder:2017mvr}.
Various calculations that combine the QCD resummation formalism of logarithmically enhanced contributions at small-\qt{} \cite{Dokshitzer:1978yd,Parisi:1979se,Collins:1984kg,Catani:2013tia}
with fixed-order perturbative results at different levels of
theoretical accuracy have been performed in
Refs.~\cite{Balazs:1997xd,Ellis:1997sc,Ellis:1997ii,Bozzi:2010xn,Banfi:2012du,Guzzi:2013aja,Catani:2015vma,Bizon:2018foh,Bizon:2019zgf}.
Analogous resummed calculations have been performed by applying 
Soft Collinear Effective Theory 
methods \cite{Becher:2010tm,Becher:2011xn,Chiu:2012ir,Ebert:2016gcn,Becher:2019bnm}
and transverse-mo\-men\-tum dependent
factorisation \cite{Collins:2011zzd,GarciaEchevarria:2011rb,Collins:2012uy,Collins:2014jpa,Scimemi:2017etj,Bertone:2019nxa,Bacchetta:2019tcu,Bacchetta:2018lna,Bozzi:2019vnl}.
Electroweak (EW)~
\cite{Dittmaier:2001ay,Baur:2004ig,Zykunov:2006yb,Arbuzov:2005dd,CarloniCalame:2006zq,Baur:2001ze,Zykunov:2005tc,CarloniCalame:2007cd,Arbuzov:2007db}
and mixed
QCD-EW~\cite{Kotikov:2007vr,Kilgore:2011pa,Dittmaier:2014qza,Bonciani:2016wya,Cieri:2018sfk,deFlorian:2018wcj,Delto:2019ewv}
radiative corrections have also been considered. 
A reliable estimate of the theoretical uncertainties 
requires various procedures, which also include variations of PDFs,
renormalisation and factorisation scales, and SM parameters.
It is thus necessary to rely on computing codes that allow fast
calculations of these variations with small numerical uncertainties.
The \dyturbo{} program, which is presented in this paper,
aims at providing
fast and numerically precise predictions of the 
Drell--Yan production cross sections, for phenomenological
applications such as QCD analyses and extraction of fundamental
parameters of the SM. The enhancement in performance over
original programs is achieved by overhauling pre-existing codes, by
factorising the differential cross section into production and
decay variables, and by introducing the usage of one-di\-men\-sion\-al and
multi-di\-men\-sion\-al numerical integration based on interpolating
functions.
The \dyturbo{} program is a reimplementation of the
\dyres~\cite{Catani:2015vma} and~\dyqt~\cite{Bozzi:2010xn} programs
for \qt{} re\-summation, and of the \dynnlo~\cite{Catani:2009sm} program 
for the finite-order perturbative QCD calculation up to NNLO.
The \dyres~\cite{Catani:2015vma} and \dyqt~\cite{Bozzi:2010xn} programs
encode the \qt{} re\-summed cross sections up to
next-to-next-to-leading-logarithmic (NNLL) accuracy by using the resummation
formalism proposed in
Refs.~\cite{Catani:2000vq,Bozzi:2005wk,Bozzi:2007pn}.
The
\Wboson+jet and \Zgamma+jet predictions at $O(\as)$ and $O(\as^2)$ 
are reimplemented from the
analytical calculations of
Refs.~\cite{Ellis:1981hk,Arnold:1988dp,Gonsalves:1989ar}, as encoded in
\dyqt{}, for the case of the triple-differential production cross
sections as a function of rapidity $y$, invariant mass $m$, and
transverse momentum \qt{} of the lepton pair, and from the MCFM
program~\cite{Campbell:2010ff}, as encoded in \dyres{} and \dynnlo,
for the 
full kinematical dependence of the decaying leptons.
Software profiling was employed to achieve code optimisation. The most
successful optimisation strategies leading to significant performance
improvement were hoisting loop-invariant expressions out of loops,
removing conditional statements from loops to allow the compiler
performing automatic loop vectorisation, and man\-u\-al loop unrolling.
The \dyturbo{} software is based on a modular C++ structure, with a few
Fortran functions wrapped and interfaced to C++. Multi-threading is
implemented with OpenMP, and through the Cuba
library by means of fork/wait system calls~\cite{Hahn:2014fua}. A flexible user
interface allows setting the parameters of the calculation through
input files and command line options. The results are provided in the
form of text files and ROOT histograms~\cite{Brun:1997pa}.
Preliminary versions of the \dyturbo{} program were used by the ATLAS
Collaboration in
Refs.~\cite{Aaboud:2017svj,Aaboud:2016zpd,ATLAS-CONF-2018-037}.
The \dyturbo{} program is publicly available~\cite{github}.
\section{Predictions with \dyturbo}
\label{sec:predictions}
\begin{figure*}
  \begin{center}
    \subfigure[]{\includegraphics[width=\columnwidth]{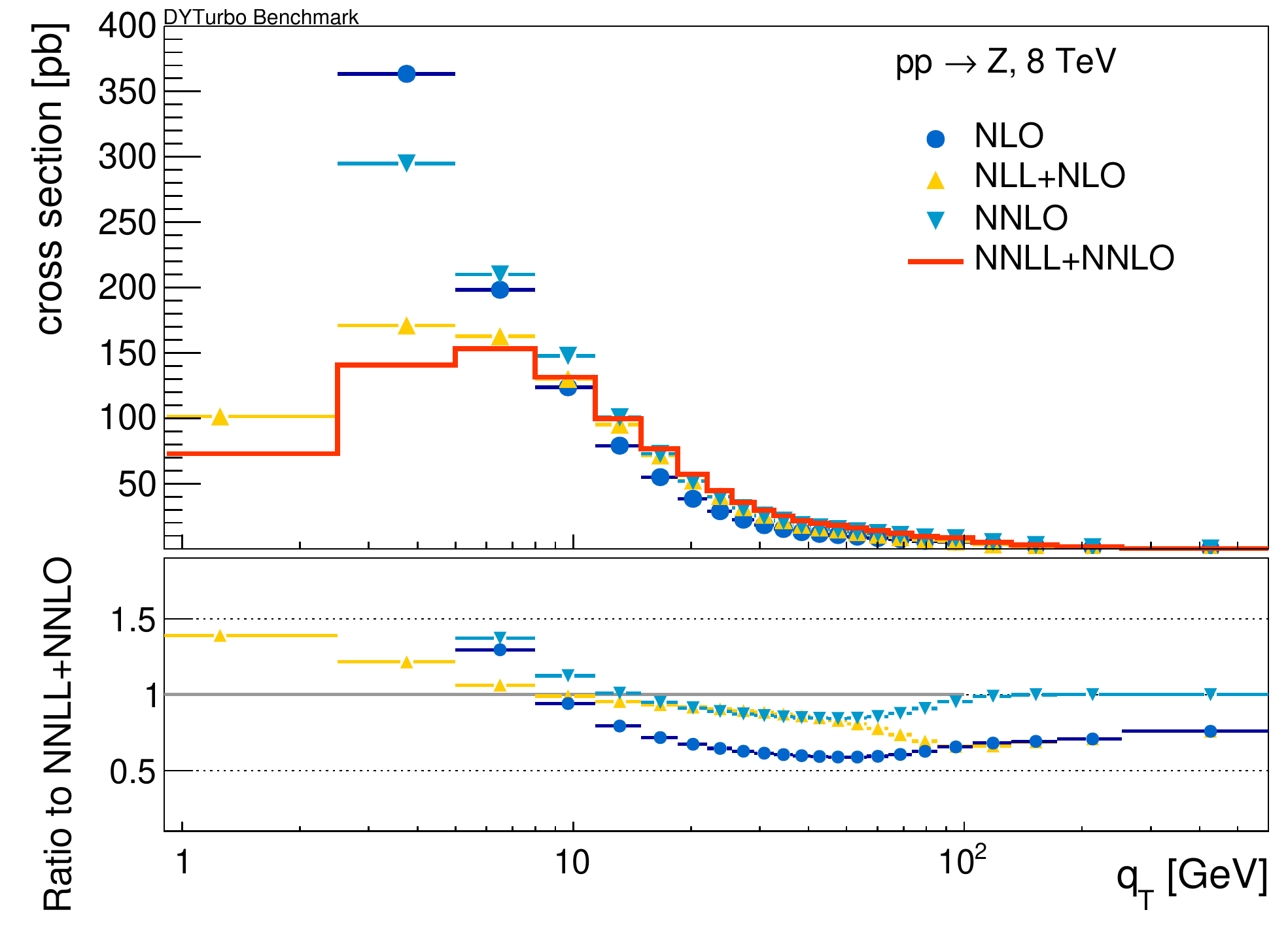}\label{fig:DYT_order}}
    \subfigure[]{\includegraphics[width=\columnwidth]{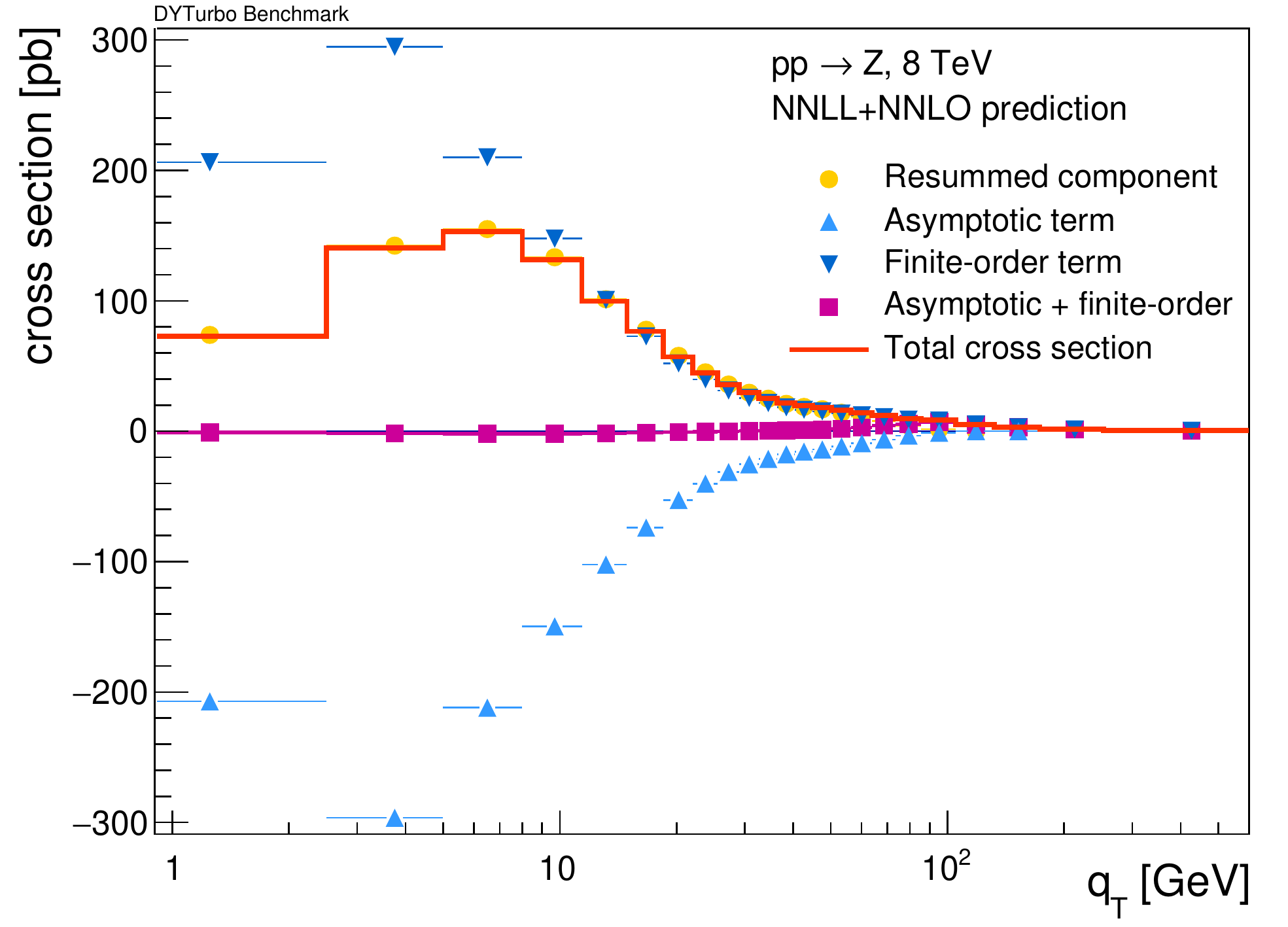}\label{fig:DYT_terms}}
    \caption{(a) 
      \dyturbo{} results for the \Zboson-boson full-lepton phase space cross
      section at $\sqrt{s} = 8$~TeV as a function of the boson transverse
      momentum \qt{} 
      at various orders of the
      calculation (NLO, NLL+NLO, NNLO, and NNLL+NNLO). The bottom
      panel shows ratios of 
      results at various 
      orders to the NNLL+NNLO
      result.
      (b) 
      \dyturbo{} results at NNLL+NNLO accuracy for the \Zboson-boson total cross
      section at $\sqrt{s} = 8$~TeV as a function of the boson transverse
      momentum and the separated contribution of various terms: resummed component,
      asymptotic term, finite-order term, sum of asymptotic and
      finite-order terms.}
\end{center}
\end{figure*}

The \dyturbo{} program provides predictions for \Wboson{} and
\Zgamma-boson (collectively denoted as \Vboson-boson) production cross
sections, fully differential in the four momenta of the decay
leptons, and inclusive over final-state QCD radiation. The cross
sections can be computed by performing the resummation of
log\-a\-rith\-mi\-cal\-ly-enhanced contributions in the small-\qt{}
region of the leptons pairs at
leading-logarithmic (LL), next-to-leading-logarithmic (NLL), and
NNLL
accuracy, and also including the
corresponding finite-order QCD contributions at next-to-leading order
(NLO) and 
NNLO.
The
log\-a\-rith\-mi\-cal\-ly-enhanced 
terms are re\-summed by using
the resummation formalism of
Ref.~\cite{Bozzi:2005wk} in impact-parameter space.
The structure of the cross section calculations is summarised in 
Eqs.~(\ref{eq:rescross_1}) and (\ref{eq:focross_1}), and we refer the reader
to the discussion in Refs.~\cite{Catani:2015vma,Bozzi:2010xn,Catani:2009sm} 
for details on the theoretical formulation.
Upon integration of final-state QCD radiation, the fully-differential
Drell--Yan cross section is described by six kinematic variables
corresponding to the momenta of the two leptons. To the purpose of
reducing the complexity of the calculation, it is useful to reorganise
the fully-differential Drell--Yan cross section by factorising the
dynamics of the boson production, and the kinematics of the boson
decay. The cross section is therefore expressed as a function of the
transverse momentum \qT, the rapidity $y$ and the invariant mass $m$
of the lepton pair, and three angular variables corresponding to the
polar angle $\theta_{\ell}$ and azimuth $\phi_{\ell}$
of the lepton decay in a given boson rest
frame 
and to
the azimuth $\phi_{V}$ of the boson in the laboratory frame.
However,
the cross section does not depend on $\phi_{V}$, since in unpolarised
hadron collisions the initial-state hadrons, i.e. the incoming beams,
are to very good approximation azimuthally symmetric. Therefore the
dependence of the cross section on $\phi_{V}$ is not considered
further.
In the following a distinction will be made between fiducial cross
sections, where kinematic requirements are applied on the final state
leptons, and total or full-lepton phase space cross sections. The
former requires the evaluation of the fivefold differential cross
sections, the latter are (\qt,$m$,$y$)-dependent triple-differential
cross sections integrated over $\cos \theta_{\ell}$ and $\phi_{\ell}$.
At NLL+NLO and NNLL+NNLO, the $\qT$-resummed cross section for
\Vboson-boson production can be written as
\begin{eqnarray}
\label{eq:rescross_1}
\textrm{d}\sigma^{\textrm{V}}_{\textrm{(N)NLL+(N)NLO}}&=&\textrm{d}\sigma^{\textrm{res}}_{\textrm{(N)NLL}}
-\textrm{d}\sigma^{\textrm{asy}}_{\textrm{(N)LO}}
+\textrm{d}\sigma^{\textrm{f.o.}}_{\textrm{(N)LO}}\, ,
\end{eqnarray}
\noindent  where $\textrm{d}\sigma^{\textrm{res}}$ is the resummed
component of the cross-section, $\textrm{d}\sigma^{\textrm{asy}}$ is
the asymptotic term that represents the fixed-order expansion of
$\textrm{d}\sigma^{\textrm{res}}$, and
$\textrm{d}\sigma^{\textrm{f.o.}}$ is the \Vboson+jet finite-order
cross section integrated over final-state QCD radiation. All the cross
sections are differential in $\qT^2$.
The resummed component $\textrm{d}\sigma^{\textrm{res}}$ is the most
important term at small \qt. The finite-order term
$\textrm{d}\sigma^{\textrm{f.o.}}$ gives the larger net contribution
at large \qt. The fixed-order expansion of the resummed component
$\textrm{d}\sigma^{\textrm{asy}}$ embodies the singular behaviour of
the finite-order term, providing 
a smooth behaviour of Eq.~(\ref{eq:rescross_1}) as
\qt{}
approaches zero. 
The two finite-order terms of Eq.~(\ref{eq:rescross_1}) and the 
finite-order factor $\mathcal{H}^{\textrm{V}}_{\textrm{(N)NLO}}$ in 
$\textrm{d}\sigma^{\textrm{res}}$ (see Eq.~(\ref{eq:rescross_2}))
are calculated up to the same power in \as.
The resummed component and its fixed-order expansion are given by\,\footnote{The convolution
with PDFs and the sum over different initial-state partonic
contributions are implied in the shorthand notation of Eqs.~(\ref{eq:rescross_2}),
(\ref{eq:resct}) and (\ref{eq:foct}). Analogously, the inverse Fourier transformation from $b$ space to \qt{} space is implied in Eq.~(\ref{eq:rescross_2}).}
\begin{eqnarray}
\label{eq:rescross_2}
\textrm{d}\sigma^{\textrm{res}}_{\textrm{(N)NLL}} &=&
\textrm{d}{\hat \sigma}^{\textrm{V}}_{\textrm{LO}}(\qt)
\times \mathcal{H}^{\textrm{V}}_{\textrm{(N)NLO}}
\times \exp\{\mathcal{G}_{\textrm{(N)NLL}}\}\\
\label{eq:resct}
\textrm{d}\sigma^{\textrm{asy}}_{\textrm{(N)LO}} &=&
\textrm{d}{\hat \sigma}^{\textrm{V}}_{\textrm{LO}}(\qt)
\times \Sigma^{\textrm{V}}(\qt/Q)_{\textrm{(N)LO}} \,,
\end{eqnarray}
where $Q$ denotes the auxiliary resummation scale \cite{Bozzi:2005wk} that is introduced in 
$\textrm{d}\sigma^{\textrm{res}}$ and, consistently, 
in $\textrm{d}\sigma^{\textrm{asy}}$.
\noindent The term $\textrm{d}{\hat \sigma}^{\textrm{V}}_{\textrm{LO}}(\qt)$ 
is the leading-order (LO) cross
section evaluated for 
non-vanishing
values of \qt{} according to a given
\qt-recoil prescription~\cite{Catani:2015vma}, namely, with values of
$\theta_{\ell}$ and $\phi_{\ell}$
that correspond
to a
chosen dilepton rest frame. The factor $\mathcal{H}^{V}$ is the
hard-collinear coefficient function. 
The term $\mathcal{G}$ is the exponent of the Sudakov form factor and
it is originally expressed as a function of the impact parameter $b$,
which is the Fourier-conjugate variable to \qt. This term embodies 
the resummation of the log\-a\-rith\-mi\-cal\-ly-enhanced 
contributions at LL, NLL or NNLL accuracy in $b$ space. 
In order to parameterise
non-perturbative QCD effects, the Sudakov form factor includes
a non-perturbative contribution, whose simplest form is a Gaussian
form factor.
The $b$ space
expression of the Sudakov form factor is then evaluated in \qt\ space
by numerically performing the (inverse) Fourier transformation.
The function $\Sigma^{\textrm{V}}(\qT/Q)$ arises from the finite-order
expansion of $\mathcal{H}^{\textrm{V}} \times \exp\{\mathcal{G}\}$,
and it matches the 
singular behaviour of $\textrm{d}\sigma^{\textrm{f.o.}}$ in the region 
$\qT \to 0$.
An additional feature of the \dyturbo{} program is the possibility of
computing finite-order cross sections at LO, NLO and NNLO without the
resummation of log\-a\-rith\-mi\-cal\-ly-enhanced contributions.
At NLO and NNLO, the finite-order cross section for \Vboson-boson
production is computed by using the $\qT$-subtraction
formalism~\cite{Catani:2007vq}, and it is expressed as the
sum of three components:
\begin{eqnarray}
  \label{eq:focross_1}
  \nonumber \textrm{d}\sigma^{\textrm{V}}_{\textrm{(N)NLO}}&=&
  \mathcal{H}^{\textrm{V}}_{\textrm{(N)NLO}}\times\textrm{d}\sigma^{\textrm{V}}_{\textrm{LO}} \\
  &&\qquad+\left[\textrm{d}\sigma^{\textrm{V+jet}}_{\textrm{(N)LO}}-\textrm{d}\sigma^{\textrm{CT}}_{\textrm{(N)LO}}\right]\, ,
\end{eqnarray}
\noindent with 
$\textrm{d}\sigma^{\textrm{CT}}_{\textrm{(N)LO}}$ given by
\begin{eqnarray}
  \label{eq:foct} \textrm{d}\sigma^{\textrm{CT}}_{\textrm{(N)LO}} &=&
  \textrm{d}\sigma^{\textrm{V}}_{\textrm{LO}} \times
  \int_{0}^{\infty}  \textrm{d}^2\qt^\prime \, \Sigma^{\textrm{V}}(\qt^\prime/m)_{\textrm{(N)LO}} \, .
\end{eqnarray}
\noindent The LO cross-section term $\textrm{d}\sigma^{\textrm{V}}_{\textrm{LO}}=\textrm{d}{\hat \sigma}^{\textrm{V}}_{\textrm{LO}}(\qt) \delta(\qt^2)$ 
is evaluated at $\qt = 0$, and
$\textrm{d}\sigma^{\textrm{V+jet}}$ is the \Vboson+jet cross section.~\footnote{More
  precisely the term $\textrm{d}\sigma^{\textrm{V+jet}}_{\textrm{(N)LO}}$
  in Eq.~(\ref{eq:focross_1}) has to be evaluated
  with $\qT > {\qT}_{\textrm{\scriptsize cut}}$, the lower integration limit on $\qt^\prime$ in Eq.~(\ref{eq:foct})
  has to be understood to be ${\qT}_{\textrm{\scriptsize cut}}$
  and the square bracket term in
  Eq.~(\ref{eq:focross_1}) has to be evaluated in the limit
  ${\qT}_{\textrm{\scriptsize cut}} \to 0$.}.
A unitarity constraint is implemented in the resummation
formalism~\cite{Bozzi:2005wk}
so as to recover exactly the finite-order
result upon integration over \qt{} of the full-lepton phase
space resummed cross section. 
The unitarity constraint leads to the following relation:
\begin{eqnarray}
  \label{eq:matching_1} \int_{0}^{\infty} \textrm{d}\qt^2 \, \textrm{d}\sigma^{\textrm{res}}_{\textrm{(N)NLL+(N)NLO}}
  &=& {\cal H}^{\textrm{V}}_{\textrm{(N)NLO}}\times\textrm{d}{\hat \sigma}^{\textrm{V}}_{\textrm{LO}}(0) \, .
\end{eqnarray}
The terms $\textrm{d}\sigma^{\textrm{res}}_{\textrm{(N)NLL}}$ and
$\textrm{d}\sigma^{\textrm{asy}}_{\textrm{(N)LO}}$ can be,
in general, multiplied by a switching function $w(\qt,m)$ above a
given \qt{} threshold, to the purpose of reducing the contribution of
the resummed calculation in the large-\qt{} region, where small-\qt{}
resummation cannot improve the accuracy of the finite-order
calculation. 
The switching function
can spoil the unitarity constraint of
Eq.~(\ref{eq:matching_1}) 
by an amount which is smaller when the chosen \qt{} threshold is
larger. The default choice in \dyturbo{} is a Gaussian switching
function, as used in \dyres.
The Drell--Yan cross section predictions are obtained by integrating
over the kinematic variables of the two leptons, and 
over
additional variables related to QCD radiation, convolutions and
integral transforms, as described in the following
Sections. The integral transformations are evaluated by means of
one-di\-men\-sion\-al quadrature rules based on interpolating functions.
The numerical integration over the other variables is performed with
two different methods. The first method is based on the \vegas{}
algorithm~\cite{PETERLEPAGE1978192} as implemented in the Cuba
library~\cite{Hahn:2004fe}. The second method employs a combination
of one-di\-men\-sion\-al and multi-di\-men\-sion\-al numerical
integrations based on interpolating functions.
The one-di\-men\-sion\-al integrations are performed by means of
Gauss--Legendre quadrature rules, with nodes and weights 
evaluated with
the Elhay--Kautsky method~\cite{Elhay:1987:AIF:35078.214351,sandia}.
The multi-di\-men\-sion\-al integrations are evaluated with the Cuhre
algorithm~\cite{GENZ1980295,Berntsen:1991:AAA:210232.210233} as
implemented in the Cuba library~\cite{Hahn:2004fe} and in the Cubature
package~\cite{cubature}, and with a tensor product of Clenshaw--Curtis
quadrature rules as implemented in the Cubature package.
The \vegas{} integration method is available for all terms of the
resummed and fixed-order calculations, and allows evaluating predictions
for any arbitrary observable, for total and fiducial cross sections.
The numerical integration based on interpolating functions is
available for all the terms in the case of total cross sections, and
for all the terms except the finite-order term at $O(\as^2)$ in the case of
fiducial cross sections. This integration method allows calculating
only the cross sections as functions of 
\qt, $m$, and $y$.
Of these two methods, the former is the most versatile, whereas the
latter allows reaching relative uncertainties in the predicted cross
sections well below $10^{-3}$ in a time frame that is significantly
shorter than that required by the \dynnlo{} and \dyres{} programs.
The 
EW
parameters $G_\textrm{F}$, $\alpha(m_Z)$, $m_W$, $m_Z$ and
$\sin^2\theta_W$ of the Drell-Yan LO cross section are set by choosing
three parameters as input, and calculating the others according to
tree-level relations. In the following the $G_\mu$ scheme is used, in
which $G_\textrm{F}$, $m_W$, $m_Z$ are set to $G_\textrm{F} =
1.1663787 \cdot 10^{-5}$~ GeV$^{-2}$, $m_W = 80.385$~GeV, $m_Z =
91.1876$~GeV, and $\sin^2\theta_W$ and 
$\alpha$
are calculated at
tree level.
The default values of the renormalisation ($\mu_R$), factorisation ($\mu_F$)
and resummation scales are fixed to $\mu_R=\mu_F=2Q=m$. 
The prescriptions necessary to obtain the resummed results
(i.e.\ the \qt-recoil prescription, the switching
function $w(\qt,m)$ and the prescription to avoid the Landau singularity) have been
chosen following Ref.~\cite{Catani:2015vma}.
Figure~\ref{fig:DYT_order} shows results for \Zboson-boson
production in proton--proton collisions at $\sqrt{s} = 8$~TeV with the
CT10nnlo set of parton density functions~\cite{Gao:2013xoa}, and
default choices of QCD scales and 
EW
parameters.
The relative contributions of 
the various terms to
the \Zboson-boson total cross section 
are illustrated in
Figure~\ref{fig:DYT_terms}. The evaluation of each term is described
in the following subsections.
\subsection{Resummed component}
\begin{figure}
  \begin{center}
    \includegraphics[width=\columnwidth]{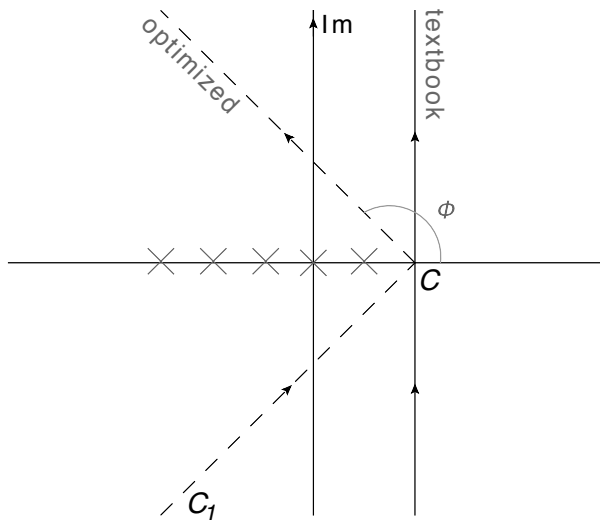}
  \end{center}
  \caption{Standard and optimised integration contours in the complex
    plane for the inverse Mellin transform. The two contours intersect
    the real axis at the point $c$, and the optimised contour is bent
    by an angle $\phi>\pi/2$ with respect to the real axis. The crosses represent the
  poles of 
  PDF parameterisation in Mellin space.}
  \label{fig:contour}
\end{figure}

The resummed component 
of the \qt-resummed cross section
(see Eq.~(\ref{eq:rescross_2}))
can be factorised as the product of the LO
cross section $\textrm{d}{\hat \sigma}^{\textrm{V}}_{\textrm{LO}}$ and the term
$\mathcal{W} = \mathcal{H}^{\textrm{V}} \times \exp\{\mathcal{G}\}$.
In these two terms, only the LO cross section depends on the
lepton angular variables, and their integration is factorised as
follows. The dependence of the cross section on $\cos \theta_{\ell}$
is $\textrm{d}\sigma(\cos \theta_{\ell}) \propto (1 + \cos^2
\theta_{\ell}) + a \cos \theta_{\ell}$,
whereas an explicit dependence on $\phi_{\ell}$ enters only in the case of
fiducial cross sections, due to the kinematic requirements
on the final-state leptons.
In the case of full-lepton phase space cross sections, the integration
over the angular variables is obtained through the following
substitutions:
\begin{eqnarray}
\label{eq:cthmom_0} 1 + \cos^2 \theta_{\ell} &\to& 
\int  \textrm{d}\Omega\,  (1+ \cos^2\theta_{\ell})= 16/3\pi\, ,\\
\label{eq:cthmom_0b} \cos \theta_{\ell} &\to& 
\int \textrm{d}\Omega\, \cos \theta_{\ell} = 0 \, ,
\end{eqnarray} 
\noindent where $\textrm{d}\Omega = \textrm{d}\cos \theta_{\ell} \, \textrm{d}\phi_\ell$.
In the more general case of fiducial cross sections, the integrals in
Eqs.~(\ref{eq:cthmom_0}) and~(\ref{eq:cthmom_0b}) are 
as follows
\begin{eqnarray}
  \label{eq:cthmom_1}
\theta_0 = \int \textrm{d}\Omega\, (1+ \cos^2\theta_{\ell})\, \Theta_K\, ,
\theta_1 = \int \textrm{d}\Omega\, \cos \theta_{\ell}\, \Theta_K\, ,
\end{eqnarray} 
\noindent where $\Theta_K$ is the acceptance function of the kinematic
requirements.
The integrals in Eq.~(\ref{eq:cthmom_1}) are evaluated by first
searching all the values of $\cos \theta_{\ell}$ corresponding to the
extremities of the region defined by the kinematic requirements at
fixed values of $\phi_\ell$. For each pair of $\cos \theta_{\ell}$
extreme values, the integrals are evaluated analytically in
$\textrm{d} \cos \theta_{\ell}$. In a second step, the integration
along $\textrm{d} \phi_\ell$ is performed by means of Gauss-Legendre
quadrature.
In the case of the full-lepton phase space cross sections, the
expressions in Eqs.~(\ref{eq:cthmom_0}) and~(\ref{eq:cthmom_0b}) do
not depend on \qt{} and $y$, which allows further simplifications. In
contrast, for the fiducial cross sections, the $\Theta_K$
acceptance function in Eq.~(\ref{eq:cthmom_1}), and so the integrals,
depend in general on \qt{} and $y$. Such a dependence varies between
different \qt-recoil prescriptions and it is of $\mathcal{O}(\qt/m)$
at small \qt.
The $\mathcal{W}$ term is expressed through the Sudakov from factor 
$\exp\{\mathcal{G}\}$ in $b$ space.
The \qt-dependent cross section is
obtained by means of a two-di\-men\-sion\-al inverse Fourier
transformation, which is expressed as 
a zeroth-order inverse Hankel
transformation
by
exploiting the azimuthal
 symmetry of the $\mathcal{W}$ function in the
transverse plane:
\begin{eqnarray}
\label{eq:bessel} \mathcal{W}(\qt,m,y) = \frac{m^2}{s} \int_0^\infty \textrm{d}b \; \frac{b}{2} \;J_0(b \qt)\;\mathcal{\tilde{W}}(b,m,y) \;,
\end{eqnarray}
\noindent where 
$\mathcal{\tilde{W}}$ is the expression of $\mathcal{W}$ in $b$ space,
$J_0(x)$ is the zeroth-order Bessel function and
$s$ is the centre--of--mass energy.
The integral transformation of Eq.(\ref{eq:bessel}) is computed by means of
a double-exponential formula for numerical
integration~\cite{OOURA1999229,OOURA1991353,DEquad}.
The convolution with PDFs is more efficiently performed by considering
double Mellin moments of the partonic functions $\mathcal{\hat{W}}_{ab}$, defined as
\begin{eqnarray}
\mathcal{\hat{W}}_{ab}^{N_1,N_2} = \int_0^1 \textrm{d}z_1 z_1^{N_1-1} \int_0^1 \textrm{d}z_2 z_2^{N_2-1} \; \mathcal{\hat{W}}_{ab}(z_1,z_2) \;,
\end{eqnarray}
\noindent where
$z_{1,2} = m /\sqrt{\hat{s}} e^{\pm\hat{y}}$, $\hat{y} = y - 1/2
\ln(x_1/x_2)$, $\hat{s} = x_1 x_2 s$, and $a,b$ denote the
initial-state partons. The Mellin moments
$\mathcal{\hat{W}}_{ab}^{N_1,N_2}$ are calculated with
\texttt{ANCONT}~\cite{Blumlein:1998if,Blumlein:2000hw}, a software library for the
analytic continuation of Mellin transformations.
The function $\mathcal{\tilde{W}}$ is then obtained by means of a
double inverse Mellin transformation:
\begin{eqnarray}
  \begin{split}
  \label{eq:mellin_inv}
&\mathcal{\tilde{W}}(b,m,y) = \left(\frac{1}{2\pi i}\right)^2 \\
  &\int_{c-i\infty}^{c+i\infty} \textrm{d}N_1 \, x_1^{-N_1} \int_{c-i\infty}^{c+i\infty} \textrm{d}N_2 \, x_2^{-N_2} \; F_a^{N_1} F_b^{N_2}  \mathcal{\hat{W}}_{ab}^{N_1,N_2} \;,
  \end{split}
\end{eqnarray}
\noindent where $x_{1,2} = m/\sqrt{s} \, e^{\pm y}$, $c$ is a real
number which lies at the right of all the poles of the integrand, and
$F_{i}^{N}$, with $i=a,b$, are Mellin moments of PDFs, $f_{i}(x)$,
defined as:
\begin{eqnarray}
\label{eq:pdfmom} F_{i}(N) = \int_0^1 \textrm{d}x \, x^{N-1} f_{i}(x) \; .
\end{eqnarray}
The integral transformation of Eq.~(\ref{eq:mellin_inv}) is computed by
means of Gauss-Legendre quadrature, and the PDFs are evolved
\cite{Bozzi:2010xn,Catani:2015vma} from the factorisation scale
$\mu_F$ to the scale $b_0/b\,$ ($b_0 = 2e^{-\gamma_E}$, and $\gamma_E$
is the Euler number) by using the program Pegasus QCD for the
evolution of PDFs in Mellin space~\cite{Vogt:2004ns}.
To perform the Mellin inversion, it is necessary to calculate the
Mellin moments $F_{i}(N)$ 
at values of $N$ along a contour of integration in the complex plane.
Parameterising the PDFs in
a simple form such as
\begin{eqnarray}
\label{eq:pdfpar} f(x) = x^\alpha(1-x)^\beta P(x) \, , 
\end{eqnarray}
\noindent where $\alpha, \beta$ are constants and $P(x)$ is a polynomial, Mellin moments
for arbitrary complex $N$ can be calculated through a simple formula
involving the $\Gamma$ function:
\begin{eqnarray}
  \label{eq:gamma}  \int_0^1 \textrm{d}x \, x^\alpha(1-x)^\beta = \frac{\Gamma(\alpha+1)\Gamma(\beta+1)}{\Gamma(\alpha+\beta+2)} \;.
\end{eqnarray}
Thanks to the analytic continuation of Eq.~(\ref{eq:gamma}) in the
region of the complex plane with $\Re(N) < 0$, when PDFs are expressed
with this form, the integration contour in Eq.~(\ref{eq:mellin_inv})
can be optimised by bending towards negative values of $\Re(N)$, as depicted schematically in
Figure~\ref{fig:contour}, allowing for a faster
convergence of the Mellin inversion integral. Such a strategy is
adopted in \dyres{} and in Refs.~\cite{PhysRevD.64.114007,Vogt:2004ns}. As a
drawback, PDFs need to be parameterised as in Eq.~(\ref{eq:pdfpar}),
or an approximation of the PDFs that follows this form has to be
evaluated, which is significantly time consuming.
In \dyturbo, the Mellin moments of PDFs are evaluated numerically, by
using Gauss-Legendre quadrature to calculate the integrals of
Eq.~(\ref{eq:pdfmom}). However these integrals can be
evaluated numerically only for $\Re(N) > 0$. As a consequence the
integration contour of the inverse Mellin transform cannot be bent
towards negative values of $\Re(N)$, and a standard 
contour along the straight line $[{c-i\infty},{c+i\infty}]$ 
is used (see Figure~\ref{fig:contour}).
This procedure results in a slower convergence of the integration in
Eq.~(\ref{eq:mellin_inv}), for which about twice as many function
evaluations are required, but it has the great advantage of allowing
usage of PDFs with arbitrary parameterisation, without requiring
knowledge of their functional form, and without requiring any time
consuming evaluation of an approximation of PDFs in the form of
Eq.~(\ref{eq:pdfpar}).
The integration over the \Vboson-boson rapidity, $y$, is factorised as
follows.
In the case of total cross sections, the values of the angular
integrals 
in Eqs.~(\ref{eq:cthmom_0})
and~(\ref{eq:cthmom_0b}) do not depend on $y$. The only dependence on
the rapidity in Eq.~(\ref{eq:mellin_inv}) is in the expression
\begin{eqnarray}
  \label{eq:rapidity}
x_1^{-N_1} x_2^{-N_2} = e^{-\ln(m/\sqrt{s}) \, (N_1+N_2)} e^{-y \, (N_1-N_2)} \,,
\end{eqnarray}
\noindent and the integrals of Eq.~(\ref{eq:rapidity}) are evaluated
analytically using the following relation:
\begin{eqnarray}
  \label{eq:yint}
\int_{y_0}^{y_1} \textrm{d}y \, e^{-y \, (N_1-N_2)} =
\frac{e^{-y_1(N_1-N_2)}-e^{-y_0(N_1-N_2)}}{N_1-N_2} \, ,
\end{eqnarray}
\noindent where $y_0$ and $y_1$ are the lower and upper $y$-bin boundaries.
In the case $N_1 = N_2$, Eq.~(\ref{eq:yint}) further simplifies to $y_1-y_0$.
When the $y$-bin boundaries are larger than the allowed kinematic
range $|y| \leq y_\textrm{max}$,
with $y_\textrm{max} = \ln (\sqrt{s}/m)$, Eq.~(\ref{eq:yint}) simplifies to $2\pi i \, \delta(N_1-N_2)$,
and the double Mellin inversion is reduced to a single
Mellin inversion~\cite{Bozzi:2007pn} by setting $N_1=N_2=N$.
In the case of fiducial cross sections, the values of $\theta_0$
and $\theta_1$ in Eq.~(\ref{eq:cthmom_1}) also depend on $y$, and the integrals
\begin{eqnarray}
  \label{eq:rap_int}
  \int_{y_0}^{y_1} \textrm{d}y \, e^{-y \, (N_1-N_2)} \, \theta_{0,1}(y) \; ,
\end{eqnarray}
\noindent are evaluated numerically for all pairs of $N_1$ and $N_2$ by means of
Gauss-Legendre quadrature.
The integration over the \Vboson-boson transverse momentum, \qt, can be
performed analytically in the case of the full-lepton phase space
cross sections, since the expressions in Eqs.~(\ref{eq:cthmom_0})
and~(\ref{eq:cthmom_0b}) do not depend on \qt, and the only term that
depends on \qt{} is $J_0(b \qt)$ of Eq.~(\ref{eq:bessel}). By using
the relation $\int \textrm{d}x \,x J_0(x) = x J_1(x)$, the integration
over \qt{} in a bin of boundaries $\qt^0$ and $\qt^1$ can be evaluated
as
\begin{eqnarray}
  \begin{split}
    &\int_{\qt^0}^{\qt^1} \textrm{d}\qt \, 2 \qt \, \mathcal{W}(\qt,m) = \\
\label{eq:qtint} \frac{m^2}{s} &\int_0^\infty {\textrm{d}b} \; \left[ \qt^1 J_1(b \qt^1) -  \qt^0 J_1(b \qt^0) \right] \;\mathcal{\tilde{W}}(b,m) \;.
\end{split}
\end{eqnarray}
Similarly to Eq.~(\ref{eq:bessel}), the integral of Eq.~(\ref{eq:qtint}) is
computed by means of a double-exponential formula for numerical
integration, and by performing two separate integrations corresponding
to the terms $J_1(b \qt^1)$ and $J_1(b \qt^0)$.
The 
information of the one-loop (two-loop) virtual correction to
the LO subprocess is contained in the $\mathcal{H}^{V}$ function.
In the computation of the fixed-order cross section of Eq.~(\ref{eq:focross_1}), the
$\mathcal{H}^{V}$ function is evaluated in $x$-space, i.e. without
performing a Mellin transformation, and the convolution with PDFs is
performed by integrating over the variables $z_{1,2} = e^{\pm\hat{y}}
\, m /\sqrt{\hat{s}}$.
The corresponding integrals are calculated with Gauss--Legendre quadrature.
\subsection{Asymptotic term and counter-term}
The asymptotic term of Eq.~(\ref{eq:resct}) and the counter-term of
Eq.~(\ref{eq:foct}) are computed using the 
function
$\Sigma^{\textrm{V}}(\qT/Q)$, which embodies the singular behaviour of
$\textrm{d}\sigma^{\textrm{f.o.}}$ in the limit $\qT \to 0$.
In the finite-order case the counter-term contributes 
at $\qt =
0$. Accordingly, the LO cross section is evaluated at $\qt = 0$ and the function
$\Sigma^{\textrm{V}}(\qT^\prime/Q)$ is integrated over the auxiliary variable 
$\qT^\prime$. 
At variance,
in the resummed case the asymptotic term is a
function of \qt{}, and the LO cross section is evaluated for nonzero
values of \qt{} according to a given \qt-recoil prescription.
As for the resummed term, 
the integration over the angular variables is factorised in the LO cross
section by using Eqs.~(\ref{eq:cthmom_0}) and~(\ref{eq:cthmom_0b}) or
Eq.~(\ref{eq:cthmom_1}).
The function $\Sigma^{\textrm{V}}(\qT/Q)$ is evaluated in $x$-space,
i.e. with\-out performing a Mellin transformation, and the convolution with
PDFs is performed by integrating over the variables $z_{1,2}$ with
Gauss--Legendre quadrature.
In the case of full-lepton phase space cross sections, the \qt{}
dependence of the asymptotic term and of the function $\Sigma^{\textrm{V}}(\qT/Q)$
is fully embodied in a set of
four functions $\tilde{I}_n(\qt/Q)$ with $n = 1,...,4$~\cite{Bozzi:2005wk}.
The integration over \qt{} of the asymptotic term is performed by
integrating the $\tilde{I}_n(\qt/Q)$ functions with Gauss--Legendre
quadrature.
In the case of fiducial cross sections, the values of $\theta_0$
and $\theta_1$ of Eq.~(\ref{eq:cthmom_1}) also depend on \qt, and the integrals
\begin{eqnarray}
  \label{eq:qt_int}
  \int_{\qt^0}^{\qt^1} \textrm{d}\qt \,2\qt\, \tilde{I}_n(\qt/Q) \, 
\theta_i(\qt) \;,~~~~~~i=0,1 \; ,
\end{eqnarray}
\noindent where $\qt^0$ and $\qt^1$ are the lower and upper \qt-bin boundaries,
are evaluated numerically by means of Gauss--Legendre quadrature.
\subsection{Finite-order term}
The real-emission corrections are 
embodied 
in the (N)LO finite-order term of
Eq.~(\ref{eq:rescross_1}) and in the $V$+jet term of
Eq.~(\ref{eq:focross_1}) for the 
resummed and fixed-order predictions, respectively. Since \dyturbo{} provides results that are
inclusive over final-state QCD radiation, the two terms are
fully equivalent~\footnote{Resummed predictions can be computed only
  inclusively with respect to final-state QCD radiation, whereas fixed-order predictions could be evaluated differentially.}.
Two independent calculations of this term are implemented. The first calculation,
which is based on the code 
MCFM~\cite{Campbell:2010ff}, 
is fully differential with respect to the lepton angular variables and the
final-state QCD radiation. 
The second calculation, which is inclusive over the lepton angles and the QCD radiation,
implements the analytic results of
Refs.~\cite{Ellis:1981hk,Arnold:1988dp,Gonsalves:1989ar}, and it relies
in part on the code taken from \dyqt{}~\cite{Bozzi:2010xn}.
The MCFM implementation of the 
lowest-order term $\textrm{d}\sigma^{\textrm{V+jet}}_{\textrm{LO}}$
can be
evaluated by using either the \vegas{} integration method or the
numerical integration based on interpolating functions.
The MCFM implementation of the 
next-order term $\textrm{d}\sigma^{\textrm{V+jet}}_{\textrm{NLO}}$
is the most
complex part of the calculation,
and it can be evaluated only with the \vegas{}
algorithm. The reason is that this NLO calculation is based on the
Catani--Seymour dipole subtraction scheme~\cite{Catani:1996vz},
in which for each point in the phase space where the real
radiation is evaluated, a set of counter-term dipoles are
computed corresponding to various different phase-space points.
As in any local subtraction procedure, the resulting integrand presents discontinuities
and it cannot be efficiently approximated by interpolating functions.
The implementation of the analytic calculation of
Refs.~\cite{Ellis:1981hk,Arnold:1988dp,Gonsalves:1989ar} yields the
triple-differential production cross sections as a function of \qt,
$m$, and $y$ of the lepton pair, and it is used only for cross sections
inclusive over the lepton decay, evaluated with numerical integration
based on interpolating functions.
\section{Tests of numerical precision}
\label{sec:closure}
\begin{figure*}
  \begin{center}
    \subfigure[]{\includegraphics[width=0.325\textwidth]{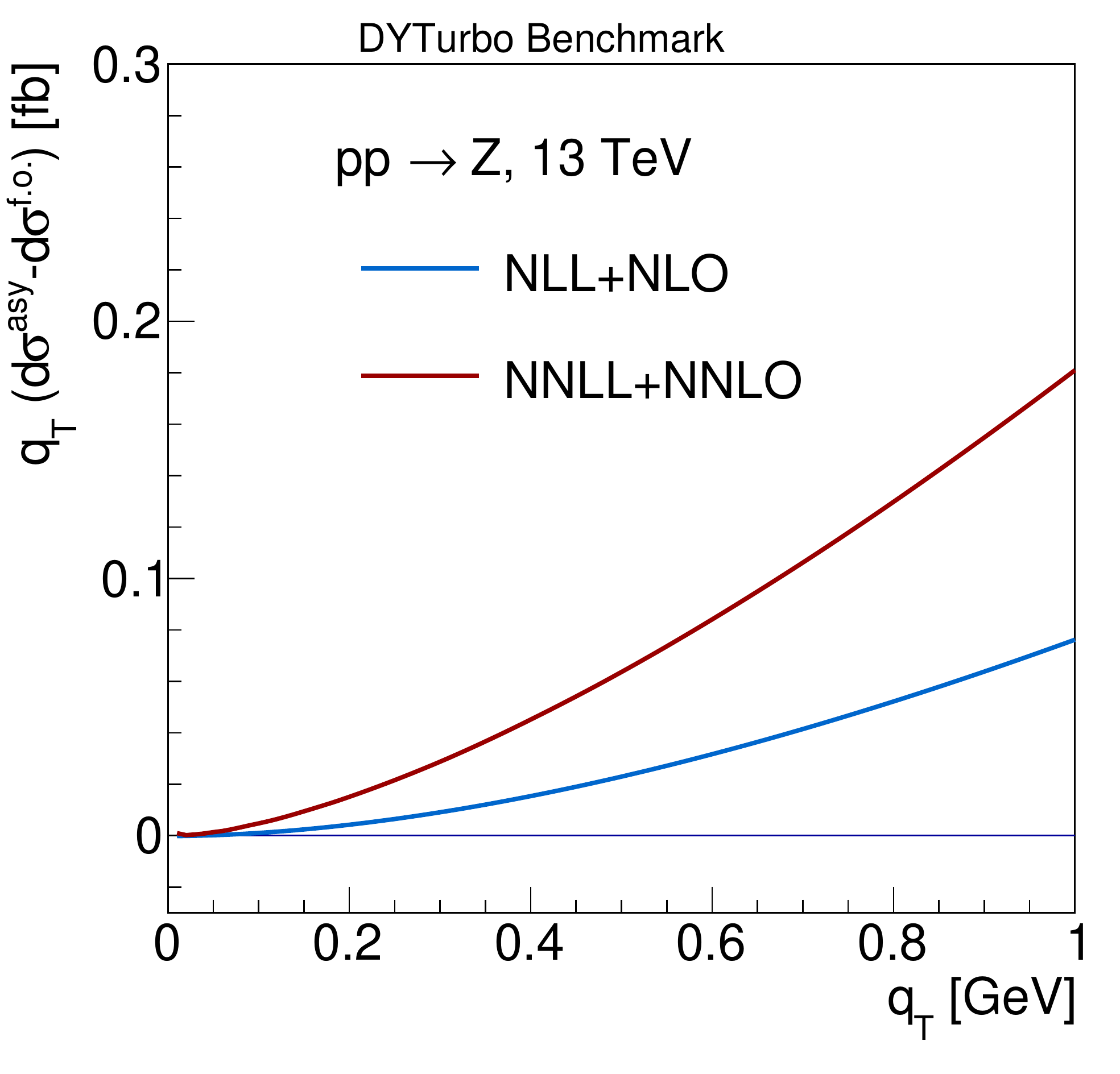}}
    \subfigure[]{\includegraphics[width=0.325\textwidth]{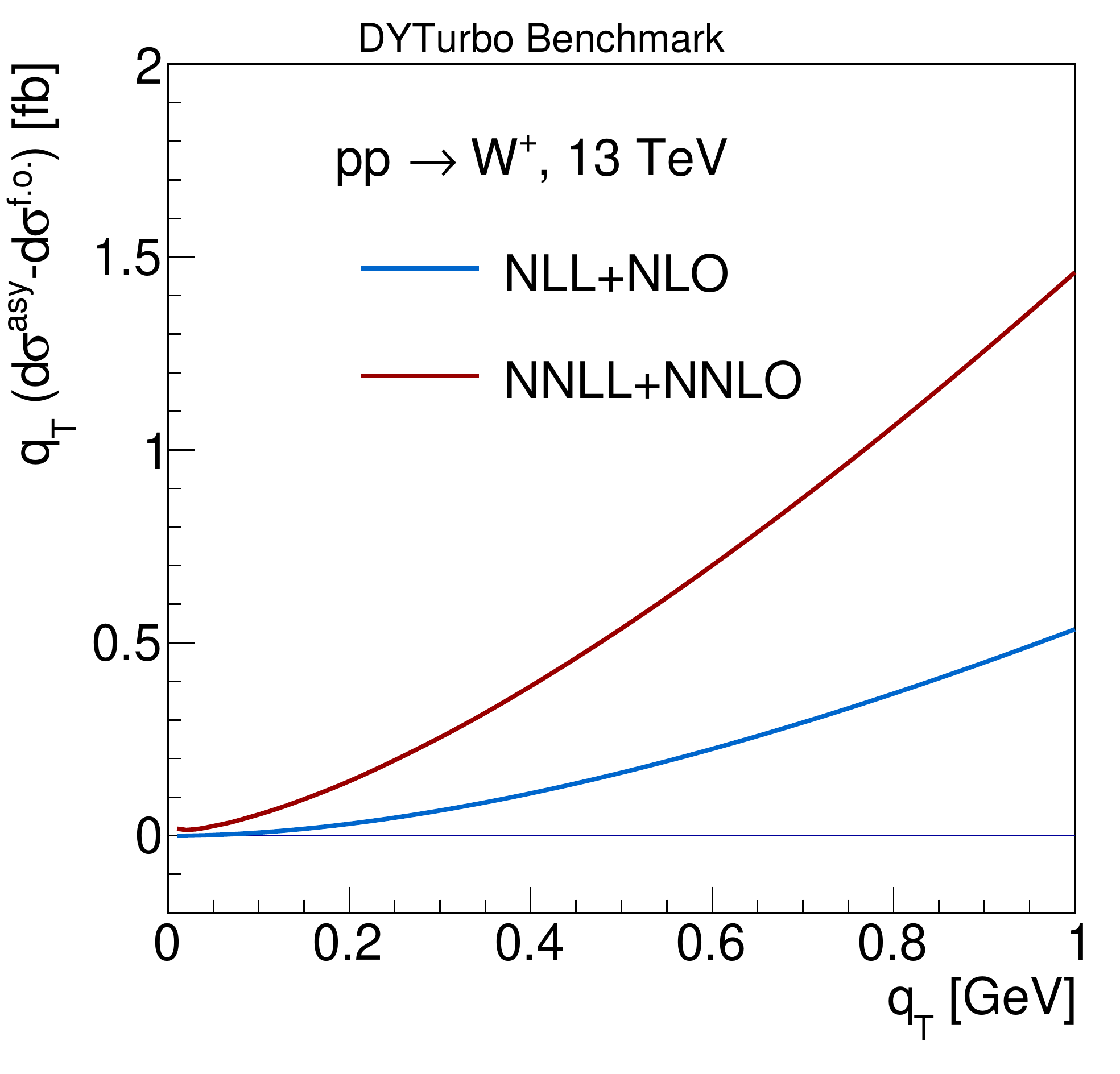}}
    \subfigure[]{\includegraphics[width=0.325\textwidth]{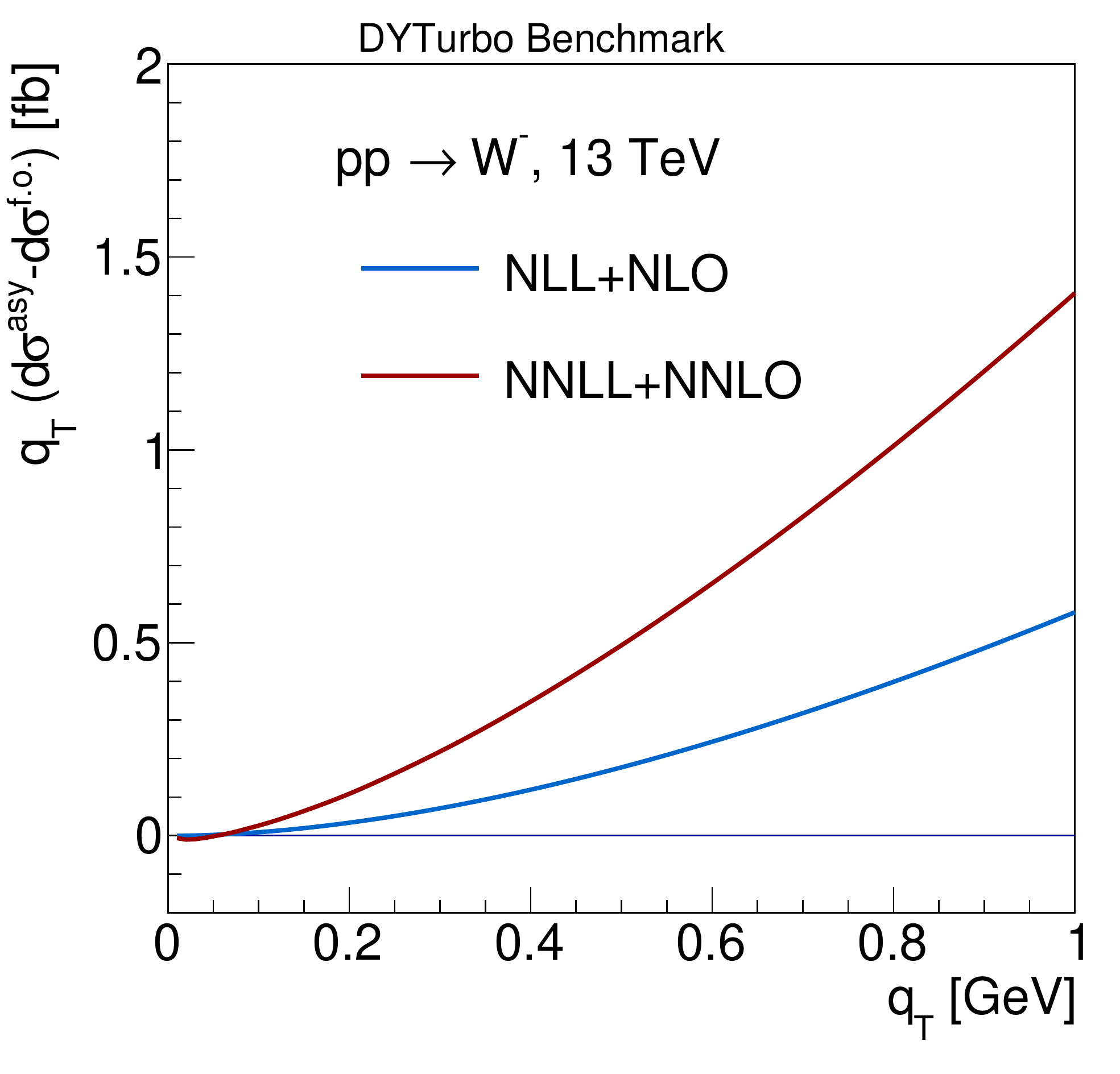}}
    \caption{Closure test of the relation $\textrm{d}\sigma^{\textrm{asy}} \sim
\textrm{d}\sigma^{\textrm{f.o.}}$ when $\qT \to 0$ for (a)
\Zgamma-boson, (b) positively-charged \Wboson-boson, and (c)
negatively-charged \Wboson-boson production at $\sqrt{s}=13$~TeV.
      \label{fig:CTVJ}}
  \end{center}
\end{figure*}

\begin{figure*}
  \begin{center}
    \subfigure[]{\includegraphics[width=0.325\textwidth]{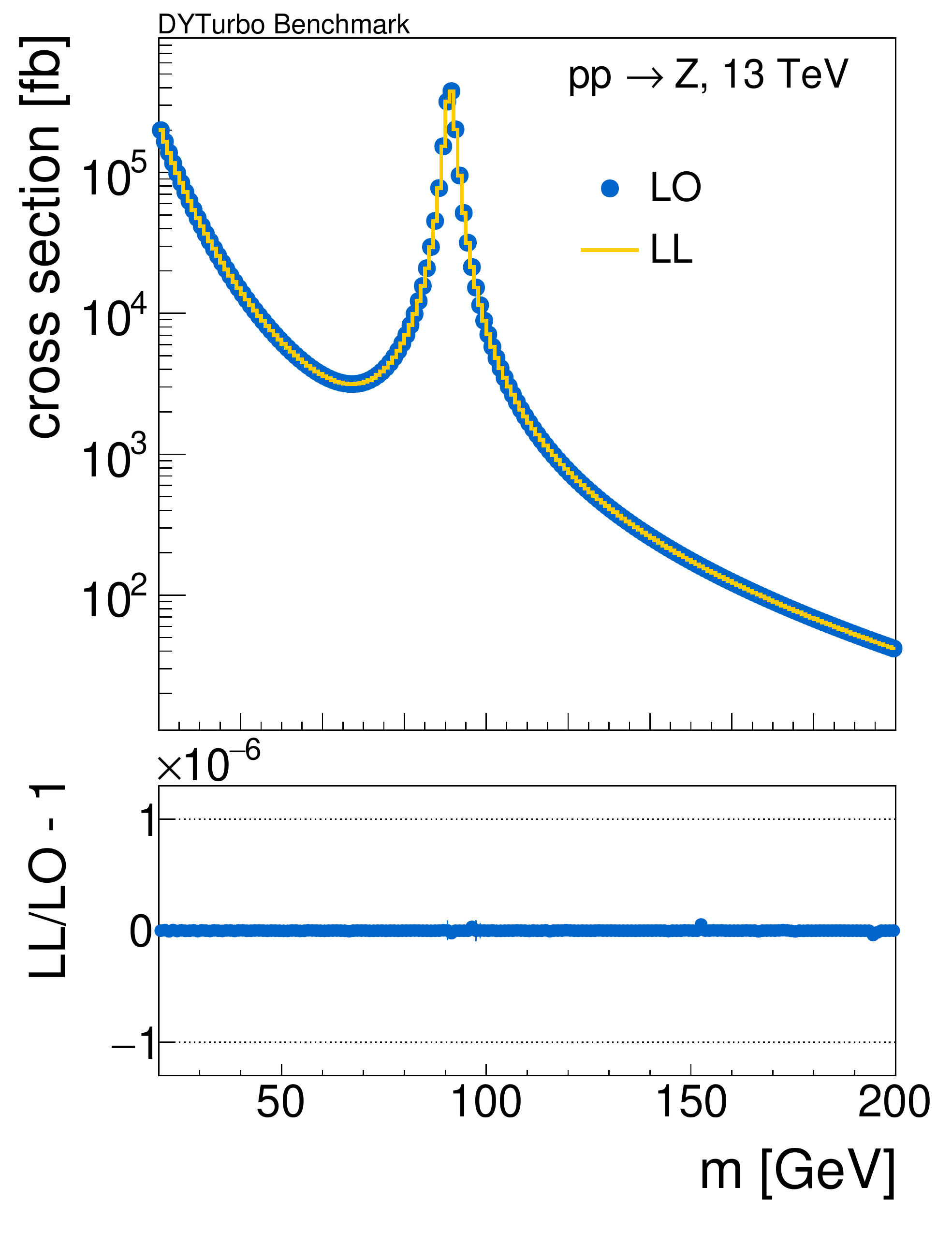}}
    \subfigure[]{\includegraphics[width=0.325\textwidth]{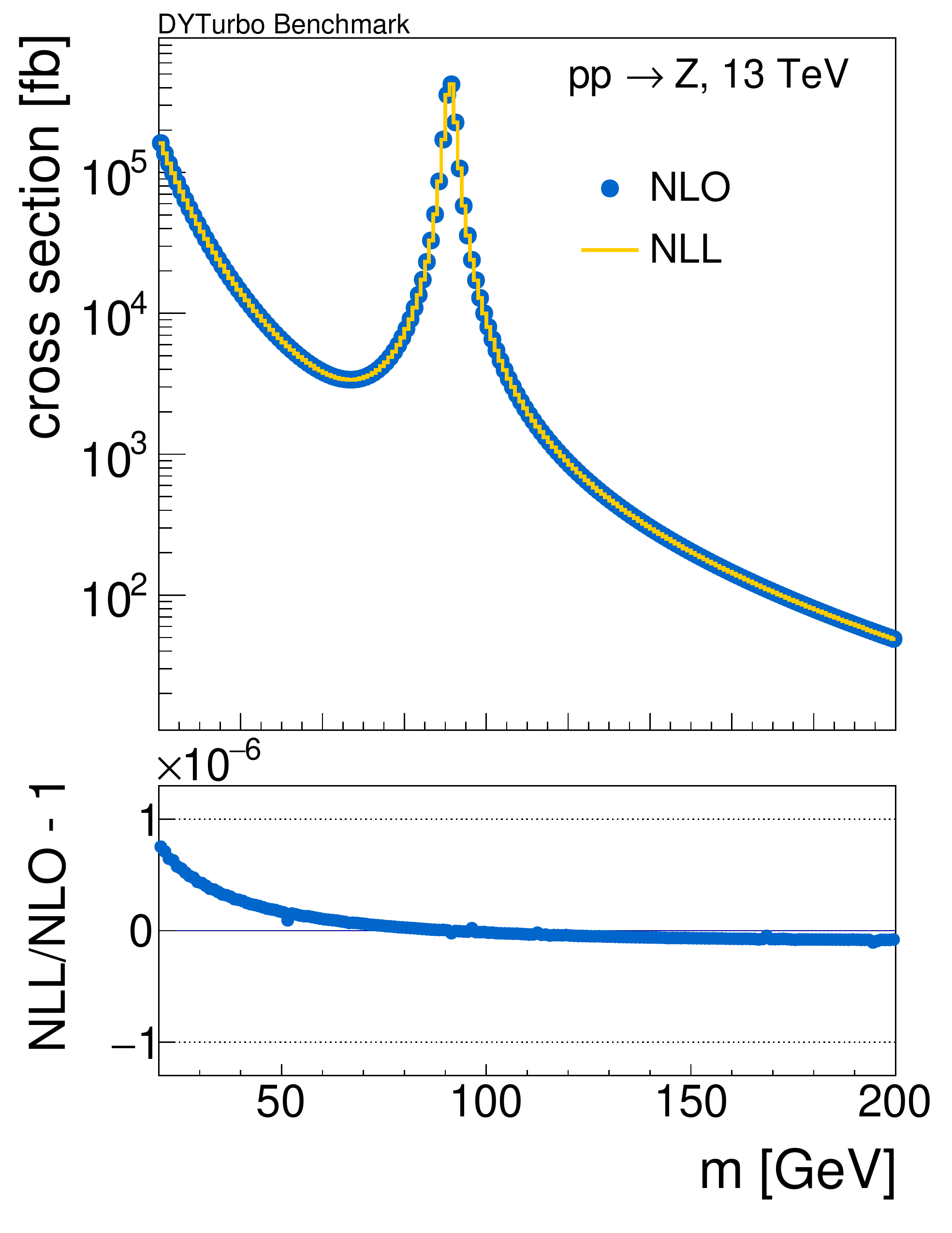}}
    \subfigure[]{\includegraphics[width=0.325\textwidth]{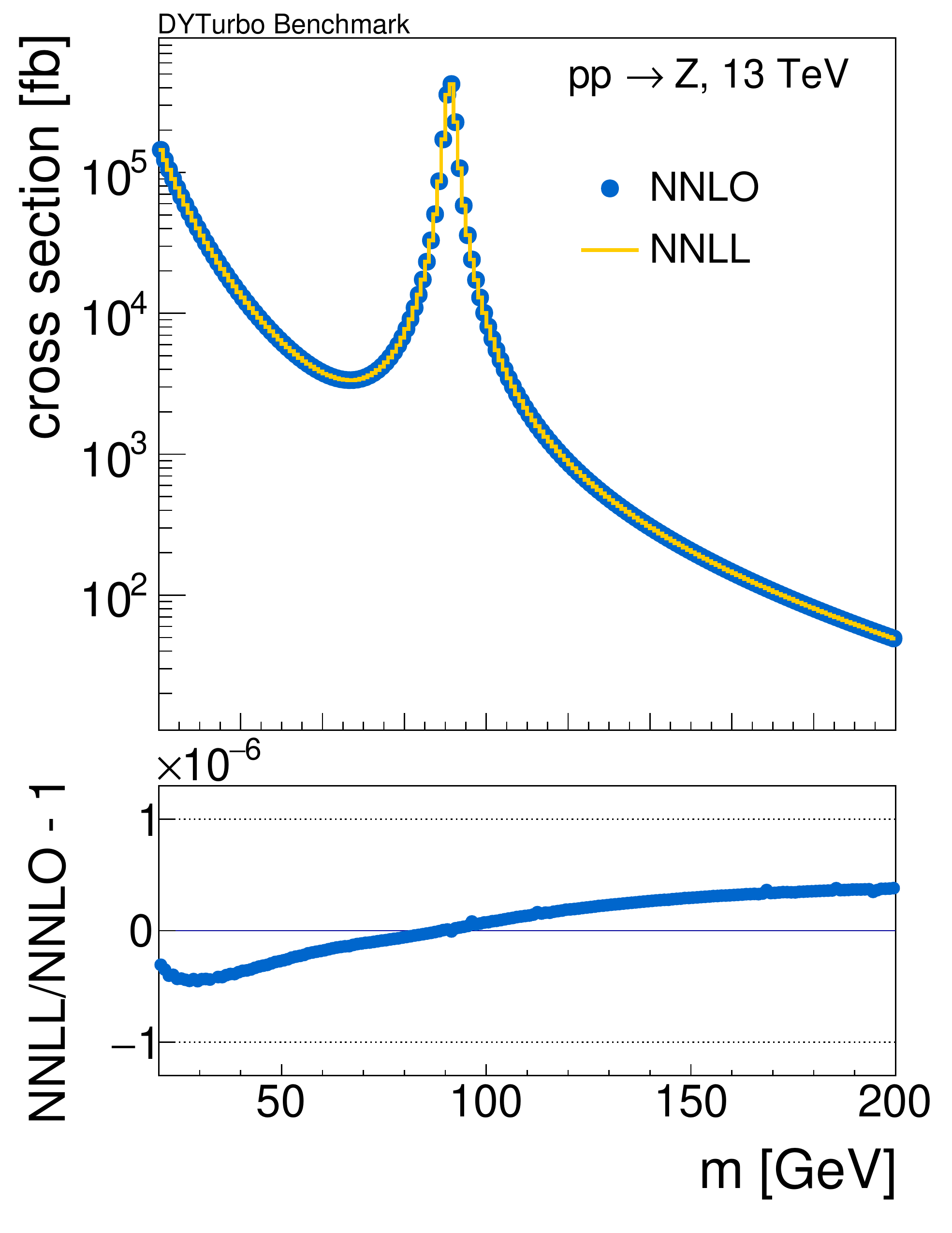}}
    \caption{Closure test of the relation
$\int_{0}^{\infty} \textrm{d}\qt^2 \, \textrm{d}\sigma^{\textrm{res}} = \mathcal{H}^{\textrm{V}}\times\textrm{d}{\hat \sigma}^{\textrm{V}}_{\textrm{LO}}$
for \Zgamma-boson production at $\sqrt{s}=13$~TeV, results at: (a) LO and LL, (b) NLO$^\textrm{virt}$ and NLL,  (c) NNLO$^\textrm{virt}$ and NNLL.
The labels NLO$^\textrm{virt}$ (NNLO$^\textrm{virt}$)
identify the contribution of
the regularised virtual corrections up to one-loop (two-loop) order.
\label{fig:matching}}
  \end{center}
\end{figure*}

\begin{figure*}[t]
\begin{center}
\includegraphics[width=\textwidth]{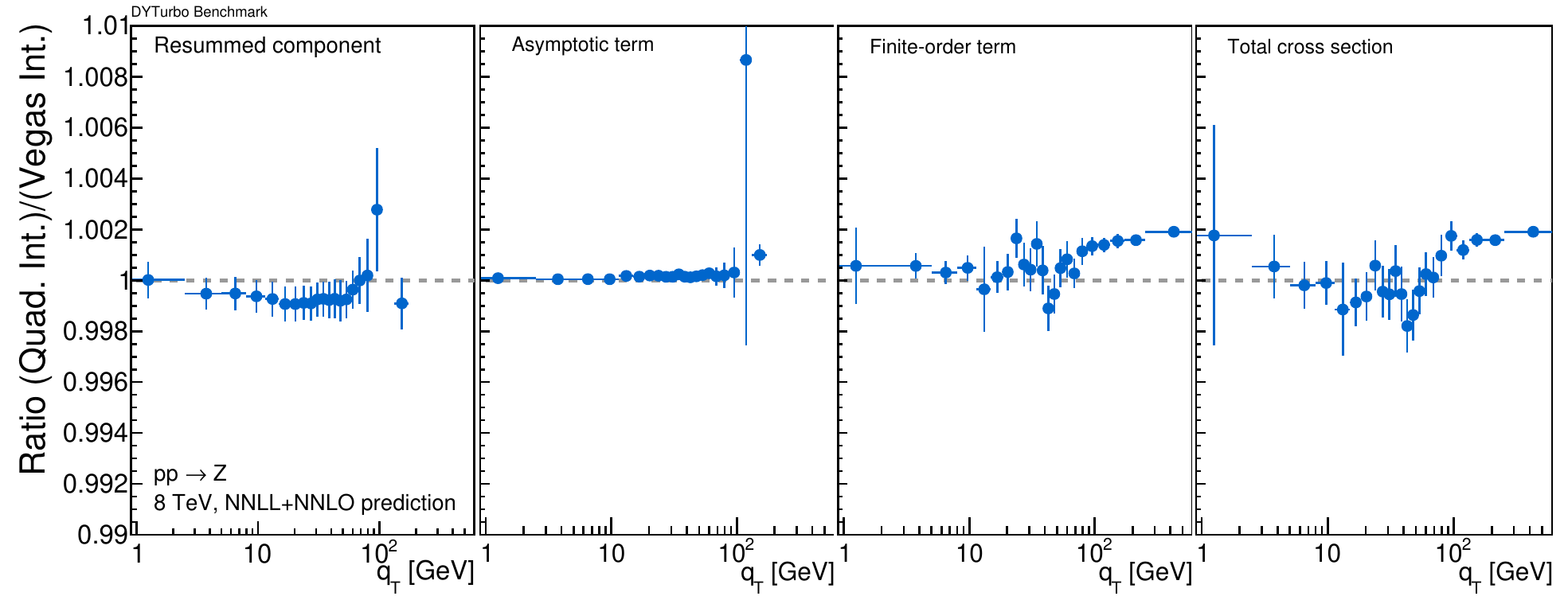}
\caption{Ratio of full-lepton phase space differential cross sections at
  $\sqrt{s}=8$~TeV as a function of \Zboson-boson transverse momentum
  as evaluated with the \quadra{} and \vegas{} integration methods of
  \dyturbo. From left to right: resummed component, asymptotic term,
  finite-order term, and total differential cross section.
\label{fig:DYTURBOvegasquadrature}}
\end{center}
\end{figure*}

In order to 
validate the numerical precision of the resummed calculation, three
closure tests are performed: the comparison of the fixed-order
expansion of the resummed component (asymptotic term) and the
finite-order term at small \qt, the comparison of the term
$\mathcal{H}^{\textrm{V}}\times\textrm{d}\sigma^{\textrm{V}}_{\textrm{LO}}$
and the resummed component upon \qt{} integration, and
comparisons of the integration methods available in \dyturbo, namely
the \vegas{} algorithm and the multi-di\-men\-sion\-al numerical
integration based on interpolating functions, referred to as \quadra{}
integration in the plots. The numerical tests of this section are
performed in full-lepton phase space, using the CT10nnlo set of
parton density functions and with default values of the QCD scales and
EW
parameters.
As discussed in Section~\ref{sec:predictions}, the function
$\textrm{d}\sigma^{\textrm{asy}}$ embodies the singular behaviour
of $\textrm{d}\sigma^{\textrm{f.o.}}$ when $\qT \to 0$, yielding the relation (see Eq.~(4) of Ref.~\cite{Bozzi:2005wk})
\begin{eqnarray}
  \label{eq:ctvj}
  \lim_{\QT\to 0}  \,\int_0^{\QT} \textrm{d}\qt^2 \left(\textrm{d}\sigma^{\textrm{\scriptsize f.o.}}
   - \textrm{d}\sigma^{\textrm{\scriptsize asy}}\right)=0\,,
 \end{eqnarray}
 or, equivalently,
 \begin{eqnarray}
   \label{eq:ctvj2}
   \lim_{\qT\to 0}  \,\qT\,\left(\textrm{d}\sigma^{\textrm{\scriptsize f.o.}}
 - \textrm{d}\sigma^{\textrm{\scriptsize asy}}\right)
   =0\,.
 \end{eqnarray}
  We note that $\textrm{d}\sigma^{\textrm{\scriptsize f.o.}}$ and $\textrm{d}\sigma^{\textrm{\scriptsize asy}}$ in Eq.~(\ref{eq:ctvj2})
   separately diverge proportionally to $\qt^{-2}$
 (modulo powers of $\log \qt$) as $\qT \to 0$.
Computing such a relation at small values of \QT{} provides a
stringent test of the numerical precision of the asymptotic and finite-order terms.
The tri\-ple-dif\-fer\-en\-tial cross sections $\textrm{d}\sigma^{\textrm{asy}}$ and
$\textrm{d}\sigma^{\textrm{f.o.}}$, as functions of \qt, $m$ and
$y$,  are evaluated at the fixed values 
$y = 0$ and $m = m_V$, with $V = W,Z$, for proton--proton collisions at $\sqrt{s} = 13$~TeV.
The result of the test is shown in Figure~\ref{fig:CTVJ} for
the NLL+NLO and NNLL+NNLO calculations. In all the cases, the relation of
Eq.~(\ref{eq:ctvj2}) is shown at values of \qt{} as low as $\qt =
0.01$~GeV.
As a second closure test, the unitarity constraint of
Eq.~(\ref{eq:matching_1}), which relates the
$\mathcal{H}^{\textrm{V}}\times\textrm{d}\sigma^{\textrm{V}}_{\textrm{LO}}$
and $\textrm{d}\sigma^{\textrm{res}}$ terms, is tested.
Computing such a relation provides a
stringent test of the numerical precision of the procedure.
The 
triple-differential cross
sections 
$\textrm{d}\sigma^{\textrm{res}}$ 
as a function of \qt, $m$, and $y$ are integrated in \qt{}
from zero to infinity, and in the range of rapidity $|y| \leq y_\textrm{max}$,
and they are compared with the
$\mathcal{H}^{\textrm{V}}\times\textrm{d}{\hat \sigma}^{\textrm{V}}_{\textrm{LO}}(0)$ double differential cross sections
as a function of $m$ and $y$, integrated in the same range of
rapidity. The switching function $w(\qt,m)$, which reduces the
contribution of the resummed calculation in the large-\qt{} region, is not
used in this test.
Figure~\ref{fig:matching} shows the result of such a comparison at LO, NLO and NNLO
for
\Zgamma-boson production in proton--proton collisions at $\sqrt{s} =
13$~TeV, for 180 equally-spaced bins of $m$ in the range
$[20,200]$~GeV.
In all cases the relation $\bigl( \int_{0}^{\infty}
\textrm{d}\qt^2 \,\textrm{d}\sigma^{\textrm{res}}\bigr) /
\bigl( \mathcal{H}^{\textrm{V}}\times\textrm{d}{\hat \sigma}^{\textrm{V}}_{\textrm{LO}}
\bigr) =1 $ 
is verified, with deviations from unity that are smaller than $10^{-6}$.
The terms $\mathcal{H}^{\textrm{V}}\times\textrm{d}\sigma^{\textrm{V}}_{\textrm{LO}}$ and
$\textrm{d}\sigma^{\textrm{res}}$ are evaluated
in $x$-space and Mellin-space, respectively. Therefore, computing such a relation also
provides a test of the numerical precision of the Mellin
inverse transformation in Eq.~(\ref{eq:mellin_inv}).
Similar level of agreement is observed by performing this closure
test as a function of the rapidity.
The 
results at NNLL+NNLO from \dyturbo{} using the numerical integration
based on interpolating functions and the \vegas{} algorithm are
compared for the \Zboson-boson differential cross section in
proton--proton collisions at $\sqrt{s}=8$~TeV. The \Zboson-boson
invariant mass range is required to be $80~{\rm GeV} < m < 100$~GeV.
Ratios of these 
results are shown in
Figure~\ref{fig:DYTURBOvegasquadrature}.
The scatter in the central values of the points is at the permille
level, except for the points at the very edges of the kinematic phase
space, and for the finite-order term at high \qt{} where deviations of
two permille are observed.
\section{Benchmark results}
\label{sec:benchmark}
This section provides benchmark results of \dyturbo{} to \dyres{} at
NNLL+NNLO for cross sections differential in \qt, and benchmark
results of \dyturbo{} to \dyres{} and \dynnlo{} for fully-integrated
fiducial cross sections at NNLL+NNLO and NNLO.
\subsection{Benchmark of \dyturbo{} to \dyres{} differential 
results
}
\label{sec:dyt_dyr}
\begin{figure*}[]
  \begin{center}
    \subfigure[]{\includegraphics[width=0.325\textwidth]{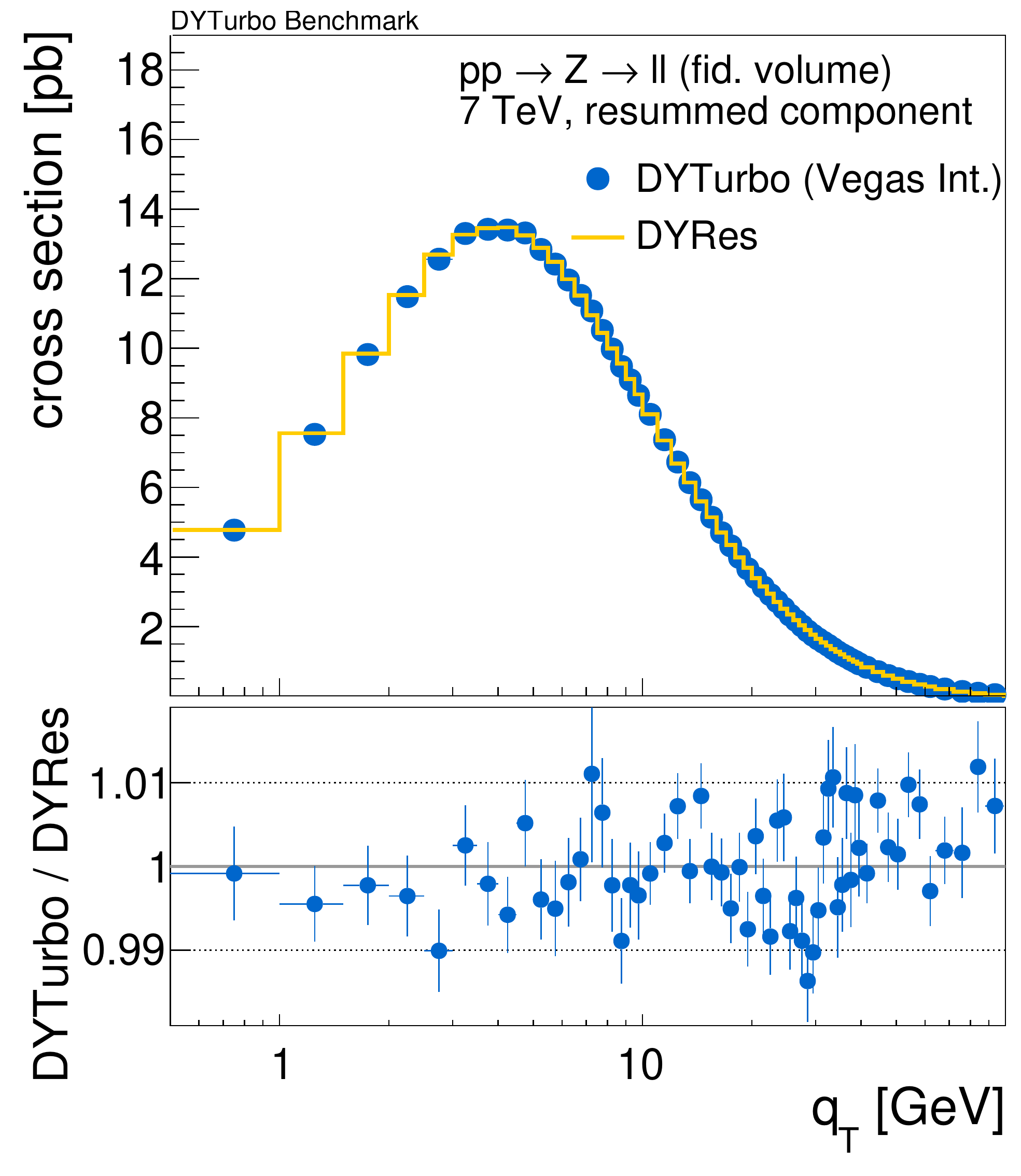}}
    \subfigure[]{\includegraphics[width=0.325\textwidth]{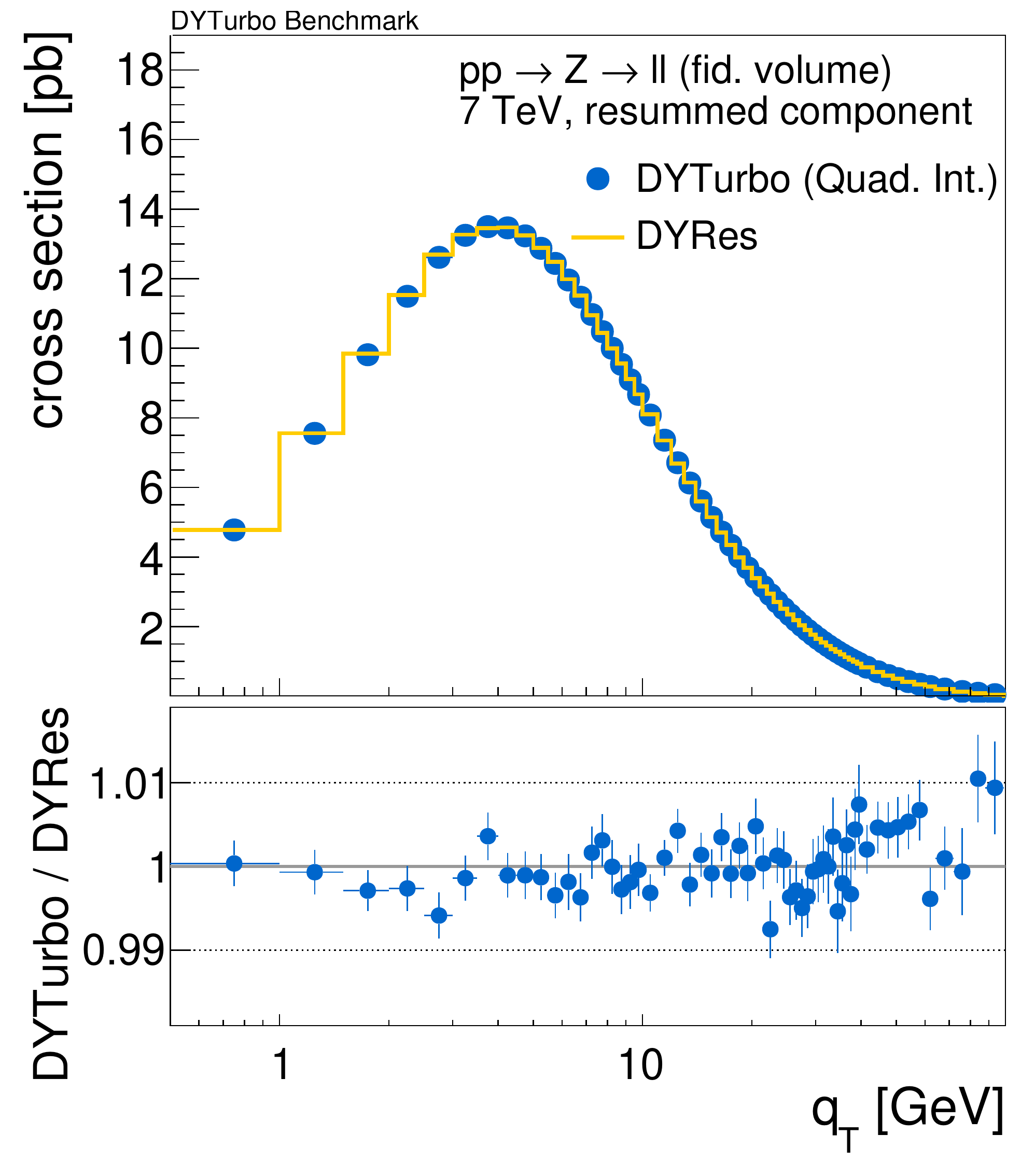}}
    \subfigure[]{\includegraphics[width=0.325\textwidth]{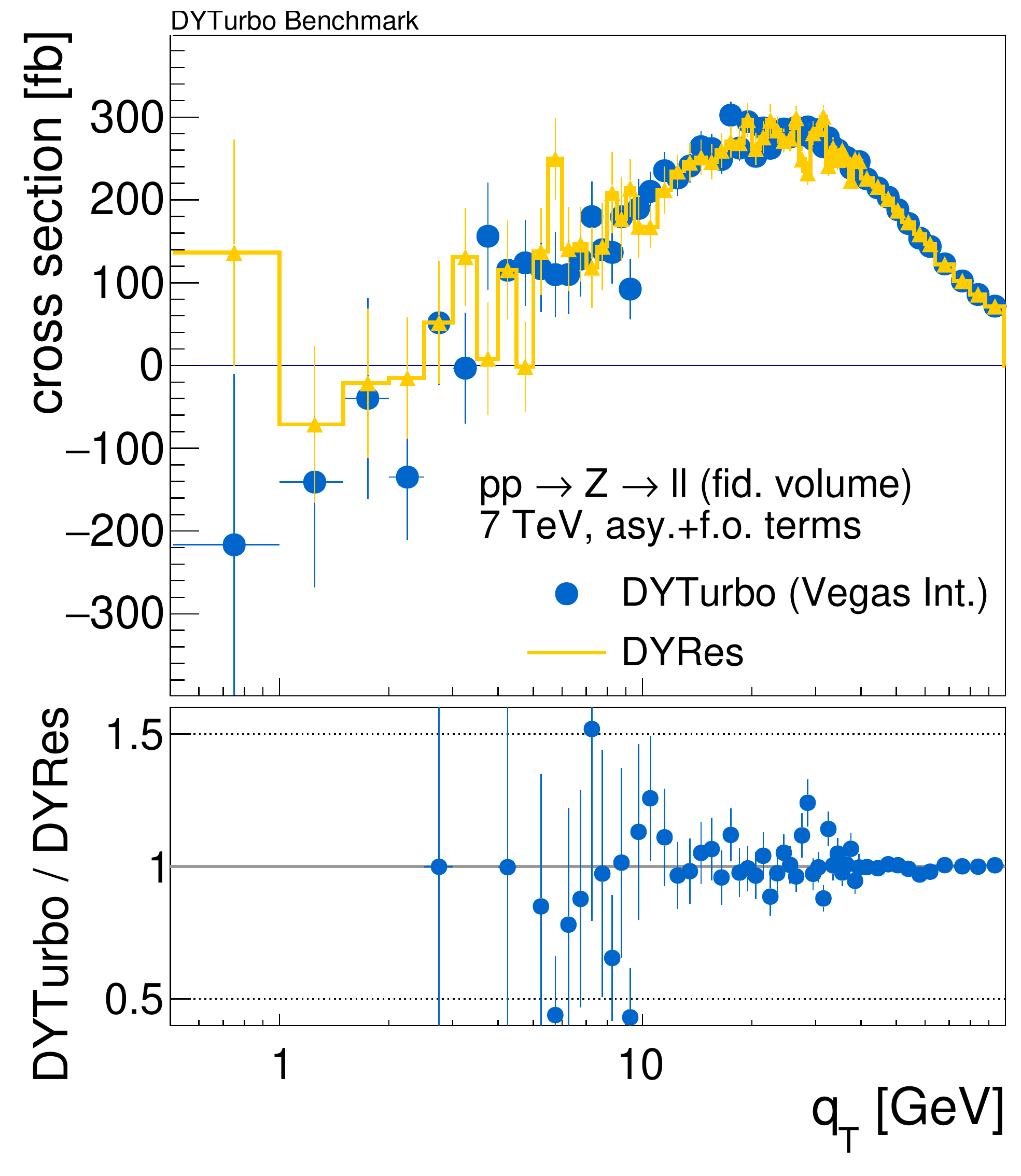}}
    \caption{Comparison of differential fiducial cross sections computed with
      \dyres{} and \dyturbo{} at $\sqrt{s} = 7$~TeV as a function of the
      \Zboson-boson transverse momentum. The \Zboson-boson fiducial
      phase space is defined by the lepton transverse momentum
      $\pT^{\ell}>20$~GeV, the lepton pseudorapidity
      $|\eta_{\ell}|<2.4$, and the invariant mass of the lepton pair 
      $66~{\rm GeV} < m < 116$~GeV. (a) Comparison of resummed component
      between \dyres{} and \dyturbo{} with \vegas{} integration. (b)
      Comparison of resummed component between \dyres{} and
      \dyturbo{} with \quadra{} integration. (c) Comparison of
      the sum of asymptotic and finite-order terms between 
      \dyres{} and \dyturbo{} with \vegas{} integration. The top
      panels show absolute cross sections, and the bottom panels show
      ratios of \dyturbo{} to \dyres{} results.
   \label{fig:ZDYRESDYTURBOcomparison}}
 \end{center}
 \end{figure*}

\begin{figure*}[]
 \begin{center}
   \subfigure[]{\includegraphics[width=0.495\textwidth]{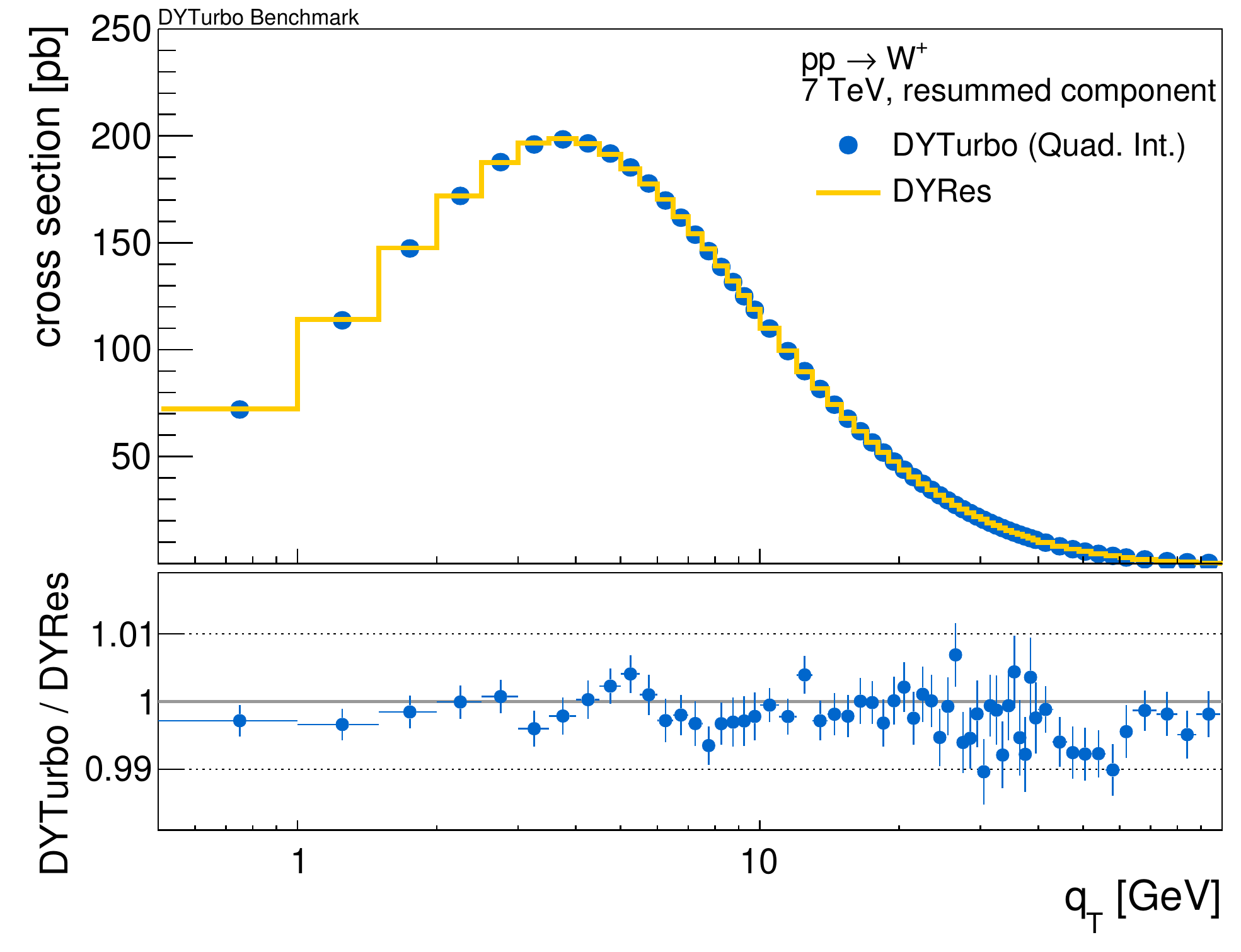}}
   \subfigure[]{\includegraphics[width=0.495\textwidth]{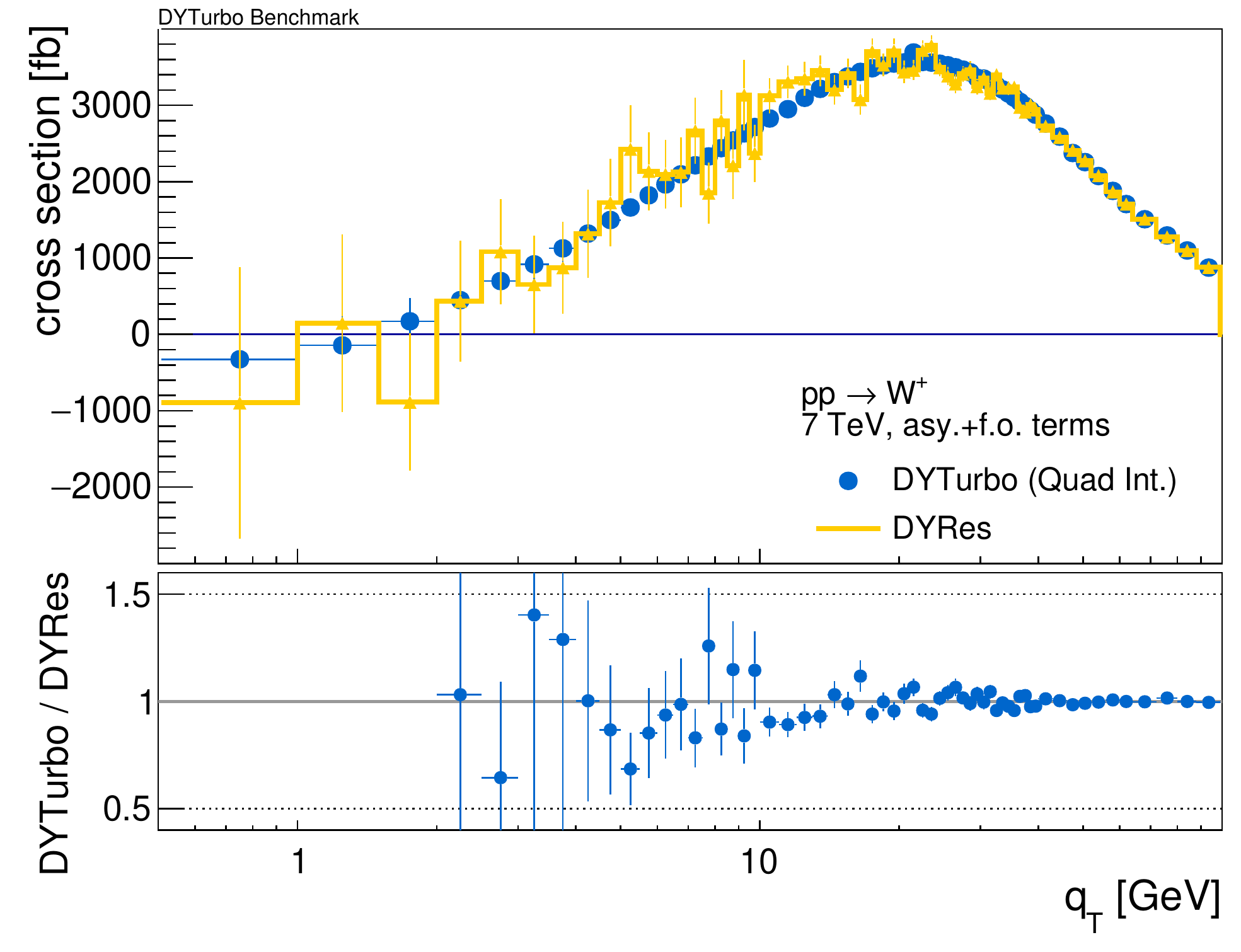}}
   \caption{Comparison of full-lepton phase space differential cross sections 
      computed with
      \dyres{} and \dyturbo{} (\quadra\ integration method)
      at $\sqrt{s} = 7$~TeV as a function of the transverse momentum of
      the positively-charged \Wboson-boson:
      (a) resummed component, (b) sum of asymptotic and finite-order 
      terms. The top panels show absolute cross sections, and the bottom
      panels show ratios of \dyturbo{} to \dyres{} results.
 \label{fig:WDYRESDYTURBOcomparison}}
 \end{center}
 \end{figure*}

\begin{figure*}[]
 \begin{center}
     \includegraphics[width=\textwidth]{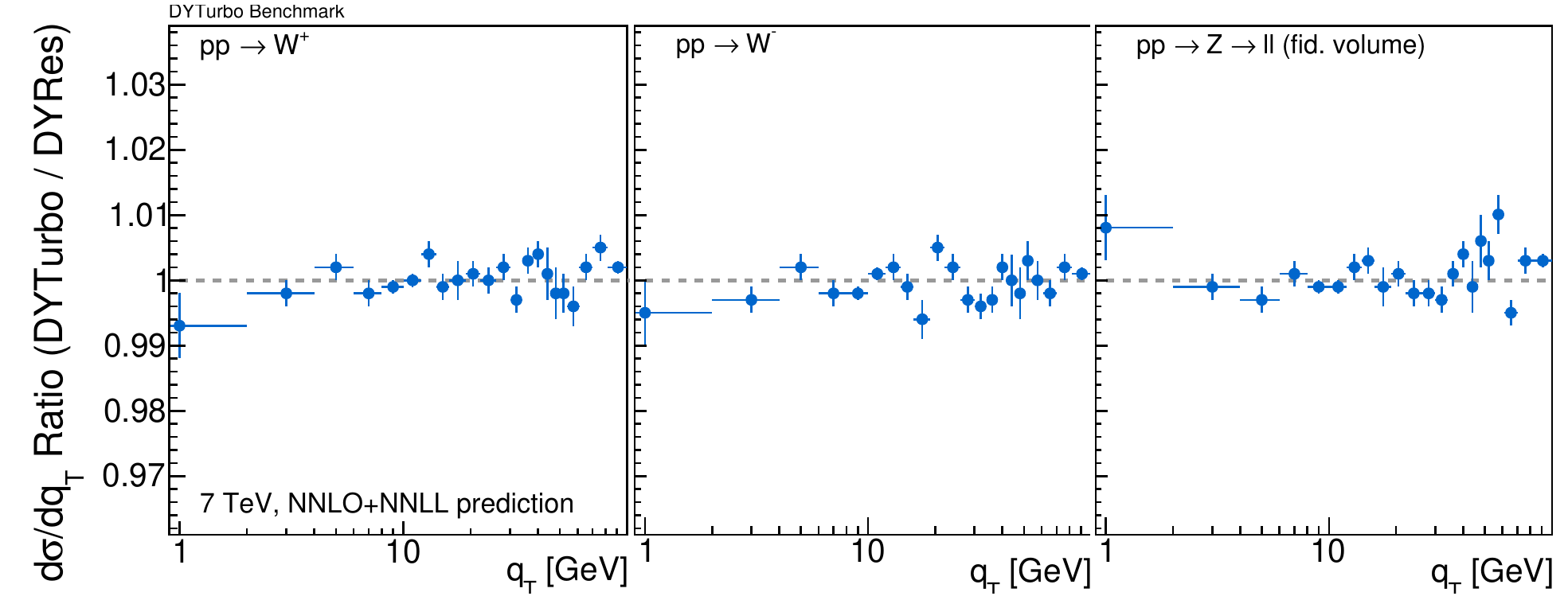}
  \caption{Comparison of \dyres{} and \dyturbo{} cross sections at
    $\sqrt{s}=7$~TeV as a
    function of the boson transverse momentum for full-lepton phase space
    $W^+$ and $W^-$ production, and fiducial \Zboson-boson production.}
  \label{fig:bench0}
  \end{center}
\end{figure*}

Predictions at NNLL+NNLO for \Zboson-boson and \Wboson-boson cross
sections in proton--proton collisions at $\sqrt{s} = 7$~TeV using the
CT10nnlo set of parton density functions were
evaluated with \dyres{}~\cite{Catani:2015vma} and compared to the
corresponding predictions in \dyturbo. The \Wboson-boson predictions
are in the full-lepton phase space, whereas the \Zboson-boson
predictions are fiducial, and match the kinematic definition of
Ref.~\cite{Aad:2014xaa}.
Careful attention is paid to exactly match in \dyturbo{} the settings
used to produce the \dyres{} predictions, such as QCD scales choice,
EW
scheme and input parameters, switching function at high \qt,
\qt-recoil prescription, and the prescription for avoiding the Landau
pole in $b$-space. All these parameters are set to the default values
of \dyres.
Figures~\ref{fig:ZDYRESDYTURBOcomparison}
and~\ref{fig:WDYRESDYTURBOcomparison} show the comparisons to
\dyres{} of \dyturbo{} 
results
for the \Zboson-boson
and positively-charged \Wboson-boson production cross sections. The
\Zboson-boson fiducial phase space is defined by the lepton transverse
momentum $\pT^{\ell}>20$~GeV, the lepton pseudorapidity
$|\eta_{\ell}|<2.4$, and invariant mass of the lepton pair in the range 
$66~{\rm GeV} < m <
116$~GeV.
All comparisons of predictions for the resummed term are validated at the better than
1\% level while the comparison of the sum of the asymptotic and finite-order
terms are validated at the $\sim$2\% level for $\qT$ above 40~GeV. In
particular, the positively-charged \Wboson-boson predictions show well
that the sum of the asymptotic and finite-order terms converges to zero at low \qt, as expected
(it is also consistent with zero for the \Zboson-boson predictions,
within the \vegas{} uncertainties, which are highly correlated
bin-to-bin).
The \dyturbo{} and \dyres{} 
results
are compared in
Figure~\ref{fig:bench0} after summing all terms.
Also in this case good agreement is observed between the two codes,
within the numerical uncertainty of the \vegas{} integration.
\subsection{Benchmark of fully-integrated cross-section 
results}
Benchmark results for fully-integrated fiducial cross section at
NNLL+NNLO from \dyres~\cite{Catani:2015vma} and at NNLO from
\dynnlo~\cite{Catani:2009sm} are shown in Table~\ref{tab:nnlo} and
compared with the corresponding results calculated with
\dyturbo\,\footnote{The NNLO results in Table~\ref{tab:nnlo} are obtained with a minimum value of $r=\qT/m$ fixed to
$r_{\textrm{cut}}=0.002$ and their corresponding numerical uncertainties do not include the systematic
uncertainty from the $r_{\textrm{cut}} \to 0$ extrapolation. A more accurate NNLO result and an
estimate of such uncertainty can be obtained by evaluating the cross section at different
values of $r_{\textrm{cut}}$ and carrying out the limit $r_{\textrm{cut}}\to 0$~\cite{Grazzini:2017mhc}.}. The
predictions are evaluated for proton--proton collisions at the
centre--of--mass energy $\sqrt{s} = 8$~TeV, and according to the
fiducial definition and QCD and 
EW
settings of
Ref.~\cite{Alioli:2016fum}. The \Zboson- and \Wboson-boson fiducial
phase space is defined by the charged lepton and neutrino transverse
momentum $\pT^{\ell,\nu}>25$~GeV, the charged lepton pseudorapidity
$|\eta_{\ell}|<2.5$, and invariant mass of the lepton pair larger than
$50$~GeV for \Zboson-boson production and larger than $1$~GeV for
\Wboson-boson production. The results for \dynnlo{} shown in the table
are taken from Table~12 of Ref.~\cite{Alioli:2016fum}. The results of
\dyturbo{} are in agreement with the results of the other programs considered in 
Ref.~\cite{Alioli:2016fum}.
Differences as large as 1\% are observed between the NNLO and the
NNLL+NNLO results, which are mostly due to recoil effects
in the lepton kinematics (the unitarity constraint of
Eq.~(\ref{eq:matching_1}) between fixed-order and resummed
calculations does not apply in the presence of lepton kinematic
cuts). An additional source of difference between
the NNLO and the NNLL+NNLO results is the inclusion of the switching function $w(\qt,m)$,
which affects the cross sections at the per mille level.

\begin{table*}
  \begin{center}
    \caption{\label{tab:nnlo} Comparison of NNLO and NNLO+NNLL cross-section results at $\sqrt{s}=8$~TeV.
      The results for \dynnlo{} are taken from
      Ref.~\cite{Alioli:2016fum}.}
        \begin{tabular}{lcccc}
      \toprule
Program                                        			& \dynnlo{} 	& \dyturbo      & \dyres{} 	& \dyturbo       \\
Order								&  NNLO   	& NNLO       	& NNLO+NNLL 	& NNLO+NNLL   	 \\
\midrule                                                                                                                         
$\sigma(pp\rightarrow W^+ \rightarrow l^+\nu$) [pb]		&$3191\pm7$	&$3176\pm3$	&$3149\pm8$   	&$3155\pm3$   	 \\
\midrule                                                                                                                         
$\sigma(pp\rightarrow W^- \rightarrow l^-\nu$) [pb]		&$2243\pm6$	& $2234\pm2$    &$2214\pm4$  	&$2213\pm2$          \\
\midrule                                                                                                                         
$\sigma(pp\rightarrow Z/\gamma^{*}\rightarrow l^+l^-$) [pb]	&$502.4\pm0.4$	& $502.8\pm0.5$	&$500.7\pm0.9$ 	&$500.5\pm0.6$ 	 \\
\bottomrule
    \end{tabular}
  \end{center}
\end{table*}

\section{Time performance}
\begin{figure*}
  \begin{center}
    \subfigure[]{\includegraphics[width=0.495\textwidth]{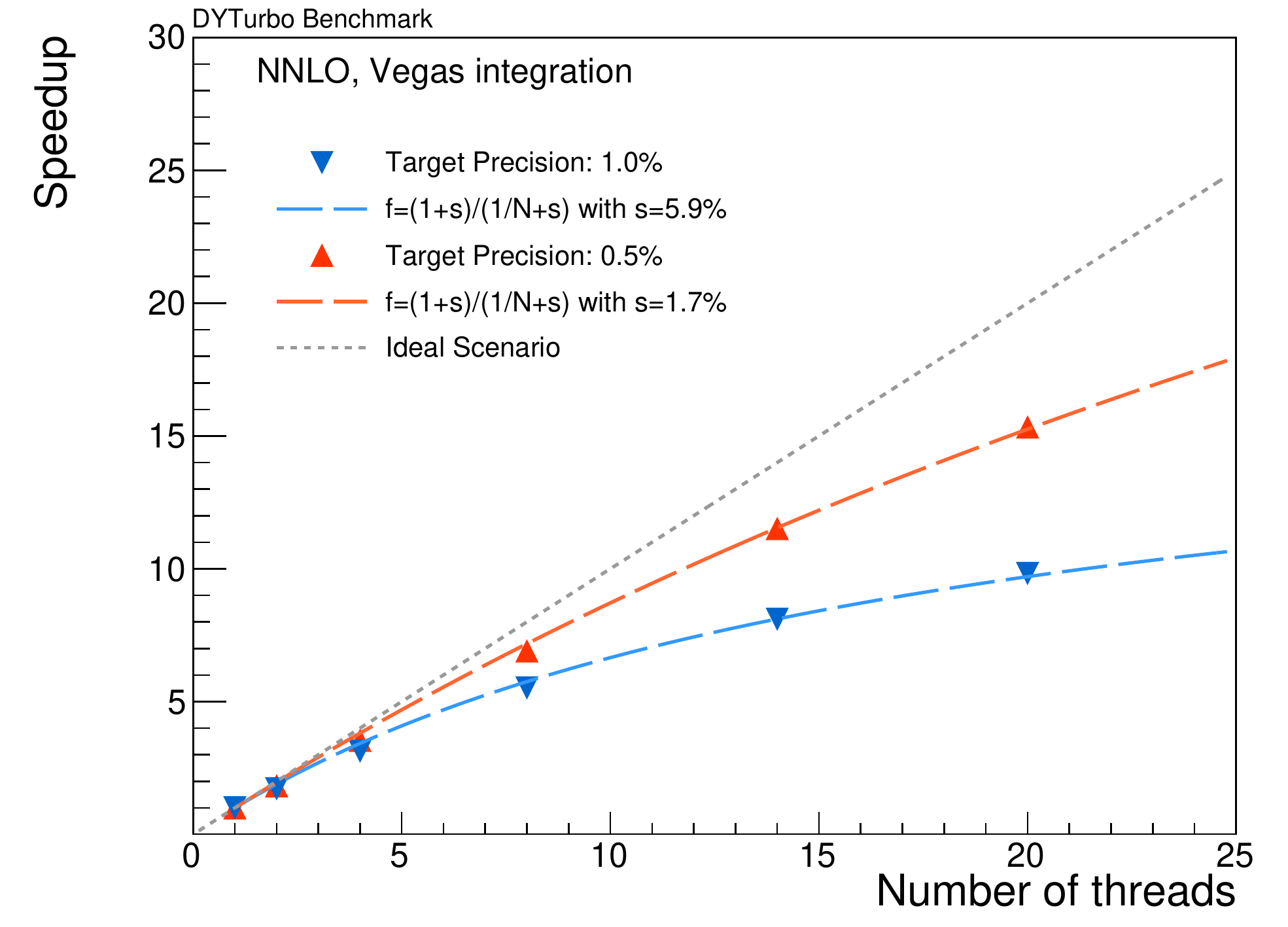}}
    \subfigure[]{\includegraphics[width=0.495\textwidth]{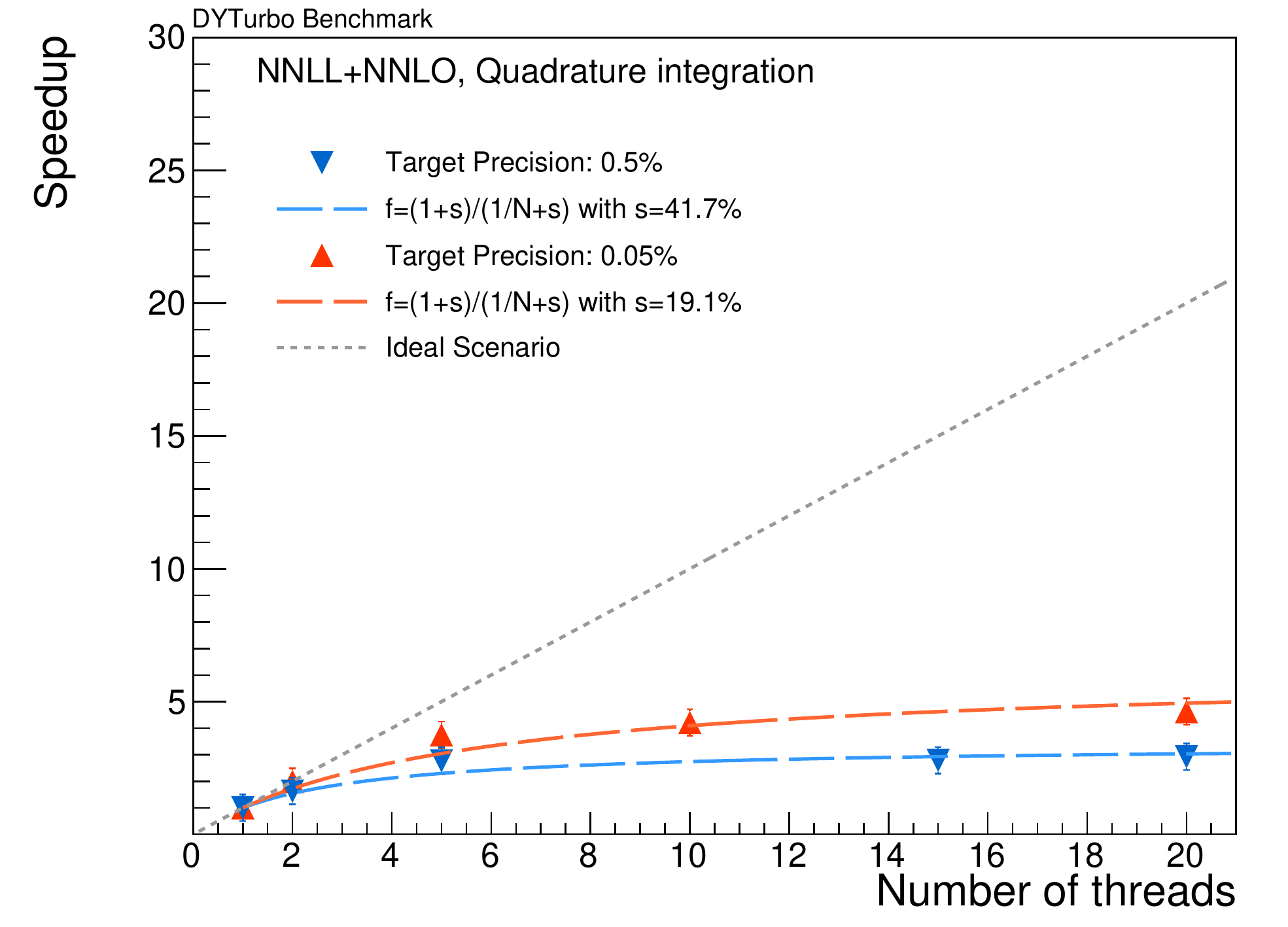}}
    \caption{Computing time as a function of the number of threads for
      \Zboson-boson production at $\sqrt{s}=8$~TeV
      with different target precision:
      (a) NNLO results with the \vegas{} integration method, and
      (b) NNLL+NNLO results with the \quadra{} integration method.
      \label{fig:SpeedBenchmark}}
  \end{center}
\end{figure*}

\begin{figure*}
  \begin{center}
    \includegraphics[width=0.6\textwidth]{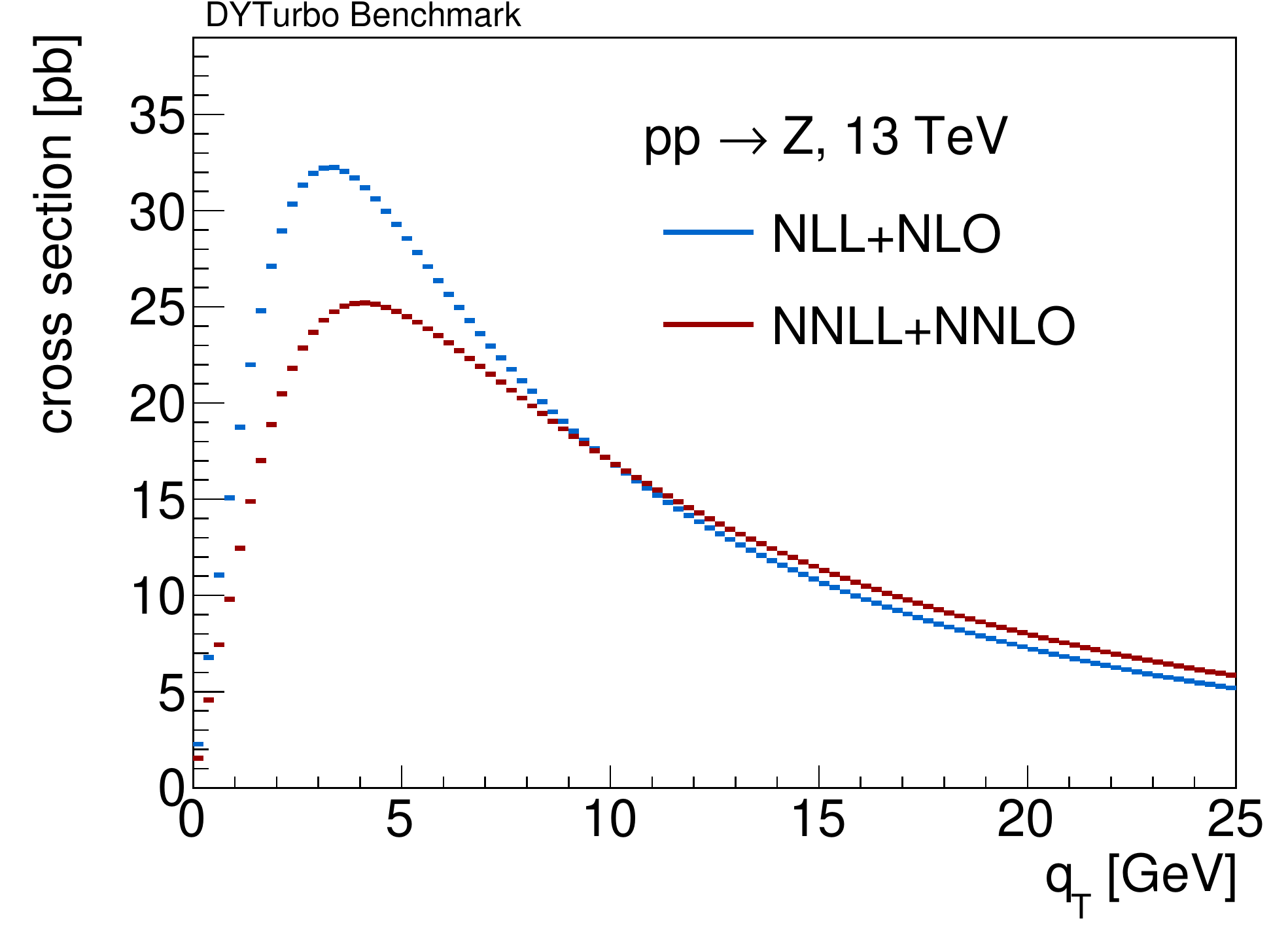}
    \caption{Example of predictions at NLL+NLO and NNLL+NNLO accuracy for
      \Zboson-boson production at $\sqrt{s}=13$~TeV.
            \label{fig:example}}
  \end{center}
\end{figure*}

In this section various tests of time performance are discussed.
The computation time requested to calculate
cross-section predictions for \dyturbo{} and \dyres{} is
compared and used to assess the performance improvement of \dyturbo.
The amount of time required to perform a calculation as a
function of threads provides a test of the scaling behaviour of the
multi-threading implementation.
The time-performance tests are run on a server mounting two AMD
Opteron 6344 CPUs with 12 cores each.
The fully-integrated fiducial cross section of \Zboson-boson
production, as defined for the 
results
shown in
Table~\ref{tab:nnlo}, is computed with \dyres{} and \dyturbo{} at NNLL+NNLO.
The \dyres{} calculation took 40~hours for an uncertainty of 0.4\
\dyturbo{} took 8~hours, yielding a factor of 5 in the improvement of the
time performance.
Figure~\ref{fig:SpeedBenchmark} shows the speedup factors for
cross-section calculations as a function of the number of threads,
where the speedup is defined as the ratio of elapsed time of the
multi-threaded calculation divided by the reference elapsed time of
the one-thread calculation. Assuming that the one-thread calculation
has a parallelisable time fraction and a non-parallelisable time
fraction, and the multithreading process has an overhead time
proportional to the number of threads $N$, the speedup curve can be
parameterised as $(1+s)/(1/N+s+o \cdot N)$. where $s$ is the ratio between
non-parallelisable and parallelisable times, and $o$ is the
overhead time per thread. The measured speedup factors are well
described by this model, with overhead times compatible with zero, and
with fractions of non-parallelisable time which are smaller when the
target precision is higher. Indeed most of the non-parallelisable time is
spent in the program initialisation, which becomes negligible for long
runs with high target precision.
We conclude this section reporting typical running times for fast
and numerically precise \dyturbo{} predictions with the numerical integration based on
interpolating functions. Figure~\ref{fig:example} shows NLL+NLO and NNLL+NNLO
predictions for the \Zboson-boson production at 13 TeV in full-lepton
phase space, integrated in the range of invariant mass $[66,116]$~GeV
and in the range of rapidity $|y| \leq y_\textrm{max} \sim 5.3$. The
predictions are computed in 100 equally-spaced
\qt{} bins from zero to 25~GeV. The predictions are evaluated with a
target in the relative numerical uncertainty of $10^{-4}$ for each
term, and using simultaneously 20 parallel threads. The computation
time required to perform the full calculation is 4~min at NLL+NLO and
3.4~hours at NNLL+NNLO. The computation of the resummed component required
6~seconds at NLL+NLO and 10~seconds at NNLL+ NNLO, the computation of the asymptotic
term required 0.2~sec\-onds at NLL+NLO and 0.7~seconds at NNLL+NNLO, the computation
of the finite-order term required 4~min at NLL+NLO and 3.4~hours at
NNLL+ NNLO. In these examples, as in all other time-per\-for\-mance tests,
the great majority of the computation time is spent to evaluate the
finite-order term. For applications as PDF fits, where very fast
predictions are required, this part of the calculation could be
computed by using APPLGRID~\cite{Carli:2010rw}.
\section{Conclusions}
The \dyturbo{} program provides fast and numerical precise predictions
of Drell--Yan processes, through a new implementation of the
\dyqt, \dyres{} and \dynnlo{} numerical codes.
The cross-section predictions include the calculation of the QCD
transverse-momentum resummation up to next-to-next-to-leading
logarithmic accuracy combined with the fixed-order results at
next-to-next-to-leading order ($\mathcal{O}(\as^2)$). They also
include the full kinematical dependence of the decaying lepton pair
with the corresponding spin correlations and the finite-width effects.
The enhancement in performance over previous programs is achieved by
code optimisation, by factorising the cross section into production
and decay variables, and with the usage of numerical integration based
on interpolating functions. The resulting cross-section predictions
are in agreement with the results of the original programs.
The great reduction of computing time for performing cross-sections
calculation opens new possibilities for the usage of Drell--Yan
processes for PDF fits, for the extraction of fundamental
parameters of the SM, such as the mass of the $W$ boson and the
weak-mixing angle, and for the estimation of background processes in
searches for physics beyond the SM.
\begin{acknowledgements}
  We thank Sven Moch for fruitful discussions on Mellin
  transformations.
  M. S. acknowledges support from the Volkswagen Foundation and the German
  Research Foundation (DFG). G. B. acknowledges support from the European
  Research Council (ERC) under the European Union’s Horizon 2020
  research and innovation program (grant agreement No. 647981,
  3DSPIN).
\end{acknowledgements}
\bibliographystyle{utphys} 
\bibliography{dyturbo}{}
\end{document}